\documentclass[11pt]{article}
\usepackage[inner=3cm,outer=3cm,bottom=2.5cm]{geometry}
\usepackage{setspace}
\usepackage{listings}
\usepackage{bm}
\usepackage[sort,comma]{natbib}
\usepackage[colorlinks=true, allcolors=blue]{hyperref}
\usepackage{changepage}
\usepackage{pdflscape}
\usepackage{booktabs}
\usepackage[normalem]{ulem}

\usepackage{graphicx}
\usepackage{float}
\usepackage{subcaption}
\usepackage{hyperref}
\usepackage{soul}
\usepackage{xcolor}
\usepackage{epsfig} 
\usepackage{times} 
\usepackage{amsmath,amsfonts,amssymb}
\usepackage{array}
\usepackage{authblk}
\usepackage[english]{babel}
\usepackage{booktabs}
\usepackage{enumerate}
\usepackage{enumitem}
\usepackage{graphicx}
\usepackage{rotating}
\usepackage{easyReview}
\usepackage{framed}

\usepackage{xfrac}  
\usepackage{mathtools} 
\usepackage{url}
\usepackage{authblk}

\title{Model--based clustering for spherical and hyper--spherical data using elliptically symmetric distributions}

\author[1]{Theodoros Perdikis}
\affil[1]{Department of Tourism Studies, University of Piraeus, Greece, \href{mailto:theoperdik@unipi.gr}{theoperdik@unipi.gr}}
\author[2]{Nader Alharbi}
\affil[2]{King Saud bin Abdulaziz University for Health Sciences, and \\
King Abdullah International Medical Research Center, Riyadh, Saudi Arabia, \href{mailto:alharbina@ksau-hs.edu.sa}{alharbina@ksau-hs.edu.sa}}
\author[3]{Michail Tsagris \thanks{Corresponding author.}}
\affil[3]{Department of Economics, University of Crete, Greece, \href{mailto:mtsagris@uoc.gr}{mtsagris@uoc.gr}}

\begin{document}
\maketitle

\begin{center}
{\bf Abstract}
\end{center}
Model--based clustering for directional data has attracted a lot of interest, but most methods utilize rotationally symmetric distributions. This paper suggests the use of elliptically symmetric distributions, namely the elliptically symmetric angular Gaussian and the spherical elliptically symmetric projected Cauchy distributions that were recently proposed in the literature for modelling spherical data. We employ the expectation--maximization algorithm and the inclusion of covariates is also examined. We also study the case of hyper--spherical data after projecting them on the sphere. Simulation studies compare the two distributions in terms of choosing the optimal number of clusters and computational cost. We use the mixtures of these two distributions to cluster two datasets on the sphere (earthquake locations) and two hyper--spherical datasets. 
\section{Introduction}
Directional data refers to multivariate data constrained to a unit norm, with the sample space represented as:
\begin{eqnarray*}
\mathbb{S}^d = \left\lbrace \mathbf{x} \in \mathbb{R}^{d+1} \bigg\vert \left|\left|\mathbf{x}\right|\right| = 1 \right\rbrace,
\end{eqnarray*}
where $\left|\left| \cdot \right|\right|$ denotes the Euclidean norm. For \(d=1\), circular data reside on a circle, while for \(d=2\), spherical data are situated on a sphere. Directional data arise in a wide range of fields, including geology \citep{chang1986}, environmental sciences \citep{heaton2014}, image analysis \citep{straub2015}, robotics \citep{bullock2014}, and space \citep{kent2016}, where direction plays a central role in understanding the underlying phenomena where directional information is essential to properly model these type of data.

Over the past decades several spherical (and hyper--spherical) distributions have been proposed. In the early $50$'s and $70$'s, \cite{fisher1953} and \cite{kendall1974} introduced two novel classes of distributions named as the von Mises--Fisher (vMF) and the projected normal distributions respectively. Recently, \cite{kato2020} proposed an extension of the Cauchy distribution to model spherical data, named as spherical Cauchy (SC) while \cite{golzy2020} considered the design of the Poisson kernel--based (PKB) distribution. However, besides the novelty and contribution of these distributions, these distributions assume rotational symmetry, which may restrict their applicability in certain scenarios. To mitigate this constraint, \cite{kent1982} introduced an elliptically symmetric distribution that relaxes the rotational symmetry assumption. This distribution constitutes a special case of the Fisher--Bingham distribution \citep{mardia1975}, and has proven valuable for modelling more sophisticated data structures. More recently, \cite{paine2018} proposed the elliptically symmetric angular Gaussian (ESAG) distribution as an alternative to the Kent distribution, possessing some nice features (closed formula for the normalizing constant, straightforward random vector generation) and with contour lines that match closely those of Kent. ESAG emerges by projecting the Gaussian distribution onto the sphere and imposing two conditions on the covariance matrix in order to avoid indeterminacy. Along similar lines, \cite{tsagris2025b} proposed the spherical elliptically symmetric projected Cauchy (SESPC) distribution, and \cite{scealy2019} proposed the scaled vMF, an elliptically symmetric distribution. These distributions, among others, have been applied to paleomagnetism \citep{scealy2019}, geology \citep{dong2024}, and in seismology to model earthquake data \citep{walsh2009,kagan2013,kagan2015}. 

\begin{figure}
\centering
\begin{tabular}{cc}
\includegraphics[scale=0.3]{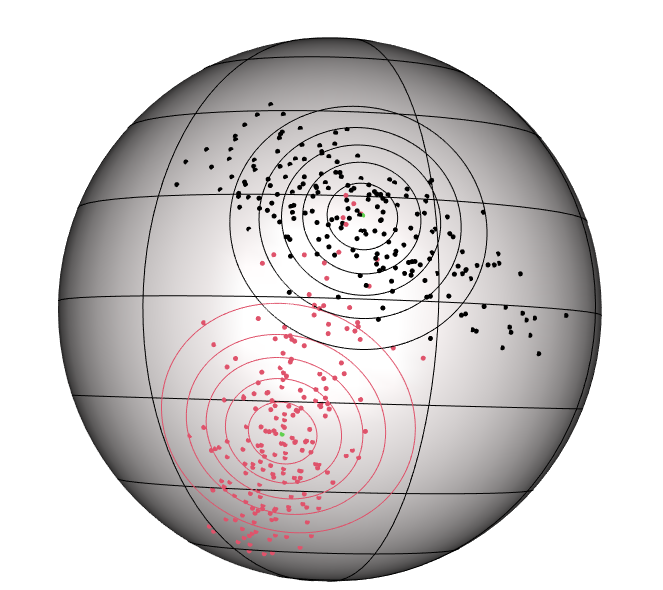} &  
\includegraphics[scale=0.3]{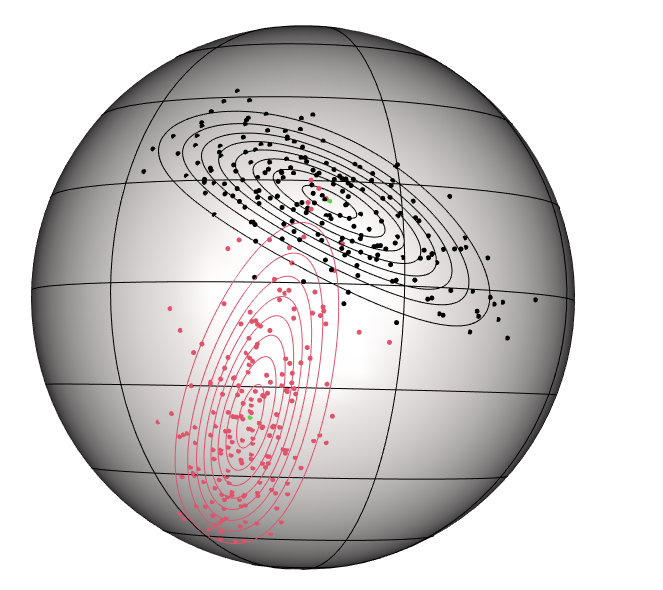} \\
vMF Model & ESAG Model \\
\includegraphics[scale=0.3]{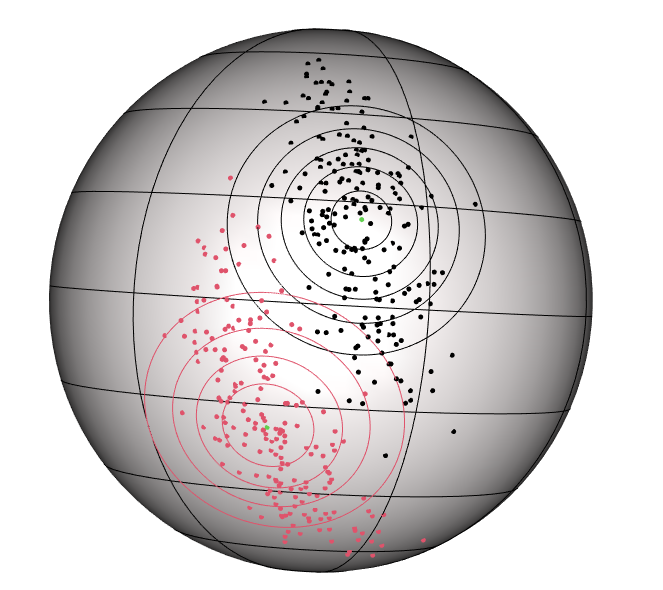} & 
\includegraphics[scale=0.3]{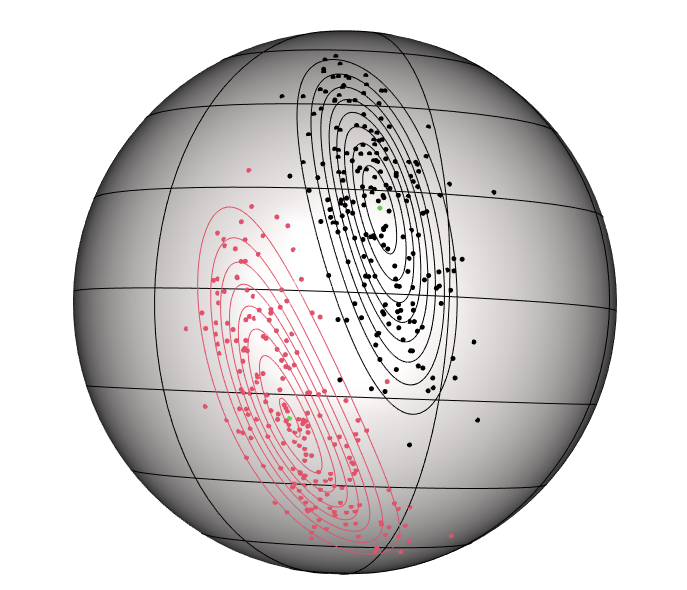} \\
vMF Model & ESAG Model \\
\end{tabular}
\caption{Contour plots for ESAG data: Modelled by the vMF (left column) and ESAG (right column).}
\label{motiv}
\end{figure}

The application of model-based clustering to directional data has attracted increasing attention, driven by the need to uncover latent groups in possibly heterogeneous spherical datasets. Under the model--based clustering framework, the observed data are modeled as a finite mixture of distributions where the  Expectation--Maximization (EM) algorithm \citep{dempster1977} is a standard approach in estimating finite mixture models \citep{mclachlan2000finite, mcnicholas2016mixture, fruhwirth2019handbook, yao2024mixture}. \cite{banerjee2005,hornik2014} proposed mixtures of vMF distributions for clustering of directional data. Ever since, many works have dealt with this problem using mixtures of VMF distributions \citep{mcgraw2006,taghia2014bayesian,gopal2014,roge2017,barbaro2021}. \cite{golzy2020} proposed mixtures of PKB distributions, and \cite{tsagris2025a} proposed mixtures of SC distributions. To the best of our knowledge, \cite{peel2001} is the only paper that proposed model--based clustering using an elliptically symmetric distribution (the Kent distribution), and more recently, \cite{dong2024} robustified the Kent mixtures model to account for outliers. Let us visualize the effect of modelling spherical data via elliptically symmetric distributions in Figure \ref{motiv}. Suppose there are two known clusters of spherical data and we model each one of them using the vMF (rotationally symmetric) and the ESAG (elliptically symmetric) distributions. Evidently, in contrast to the two vMF distributions, the two ESAG distributions have fitted the two clusters adequately. 

Although model--based clustering for directional data has received attention by the research community in recent years, the vast majority of existing approaches rely on rotationally symmetric distributions, such as the vMF, SC, and PKB. As illustrated in Figure~\ref{motiv}, rotationally symmetric distributions impose isotropic contours on the sphere, which may fail to  capture 
the true shape of clusters that exhibit elliptical structure. This limitation motivates the need for mixture models based on elliptically symmetric distributions, which allow for a more flexible covariance structure and, consequently, a more faithful representation of the underlying cluster 
geometry. Furthermore, while most existing methods are restricted to spherical data ($d=3$), many real--world applications involve data residing on the hyper--sphere ($d>3$), a setting that has received comparatively less attention in the model--based clustering literature. 

The present paper aims to fill both gaps by proposing and evaluating mixture models for both spherical and hyper--spherical data, relying on elliptically symmetry while retaining tractable parametric forms. We propose model--based clustering using mixtures of ESAG and SESPC distributions for spherical and hyper--spherical data. We employ the EM algorithm to fit the mixtures with a fixed number of components $K$, and utilize the integrated completed likelihood (ICL) criterion \citep{biernacki2000assessing} to choose the appropriate number of $K$. We compare, via simulation studies, the two distributions in terms of number of clusters, selected according to ICL, and quality of the clusters, measured by the adjusted Rand index (ARI), but also in terms of computational cost. When it comes to data that reside on the hyper--sphere we project them onto the sphere and then employ the same mixture models. We use these mixture models to cluster the locations of earthquake locations from two regions (spherical), and wine quality and wholesales data (hyper--spherical).  

The remainder of this paper is organised as follows. In Section \ref{sec:general}, some preliminaries regarding the properties and parametrization  of the two distributions under study are presented. In Section \ref{sec:em}, a  presentation  of the EM algorithm being  considered to estimate the mixture models is illustrated with and without the inclusion of covariates. Furthermore, Section \ref{sec:simulations}, contains simulation studies offering an inter comparison of the two proposed mixture models and Section \ref{sec:real} illustrates their performance modelling the aforementioned (hyper--)spherical datasets. Finally, in Section \ref{con}, some concluding remarks and
suggestions for future research are discussed.

\section{Theoretical background of the distributions}
\label{sec:general}
\subsection{The elliptically symmetric angular Gaussian distribution}
\label{sec:ESAG:general}

\cite{paine2018} proposed a new family of symmetrical distributions for spherical  data. In fact, the ESAG distribution arises from projecting a multivariate normal on the sphere. Let ${\bf X}$ be a $d$--dimensional random variable  from a $d$--variate  Normal distribution with covariance matrix $\bm{V}$ and mean vector $\bm{\mu}$ which is projected on the circle/sphere/hyper--sphere, ${\bf Y}={\bf X}/r$, where $r=\left|\left|{\bf X}\right|\right|$. The marginal distribution of ${\bf Y}$, denoted by angular Gaussian distribution (AG), arises by integrating out $r$ 

\begin{equation} 
f({\bf y})=\int_0^{\infty}r^{d-1}f(r{\bf y})dr,
\label{esag_pdf}
\end{equation}
resulting in
\begin{equation*}
f_\text{AG}(\bm{y}) = \frac{C_d}{|\bm{V}|^{1/2}}\frac{1}{B^{d/2}} \exp \left [\frac{1}{2}\left(\frac{A^2}{B}-\Gamma^2\right)\right] \mathcal{M}_{d-1}\left(\frac{A}{B^{1/2}}\right),
\end{equation*}
where $C_d=1/(2\pi)^{(d-1)/2}$,
\begin{equation*} 
A =\bm{y}^\top\bm{V}^{-1}\bm{\mu}, \  \ 
B =\bm{y}^\top\bm{V}^{-1}\bm{y}, \ \ \text{and} \ \  
\Gamma^2 =\bm{\mu}^\top\bm{V}^{-1}\bm{\mu},
\end{equation*}
and, for real $\alpha$,
\begin{align*}
\mathcal{M}_{d-1}(\alpha) =\int_{u=0}^\infty u^{d-1} \frac{1}{(2\pi)^{1/2}}\exp \left
\{-(u-\alpha)^2/2 \right \} \mathrm{d}u.
\label{trunc}
\end{align*}

\cite{paine2018} showed that:
\[ \mathcal{M}_0(\alpha)=\Phi(\alpha),  \hskip 0.1truein \mathcal{M}_1(\alpha)=\alpha \Phi(\alpha) +\phi(\alpha) \nonumber, \hskip 0.1truein \mathcal{M}_{2}(\alpha)=(1+\alpha^2)\Phi(\alpha)+\alpha \phi(\alpha),\]
where $\phi(\cdot)$ and $\Phi(\cdot)$ correspond to the p.d.f and c.d.f of the standard normal distribution, respectively. \cite{paine2018} extended (\ref{esag_pdf}) into a general family, termed elliptically symmetric
angular Gaussian distribution (ESAG), by imposing the following two conditions
\begin{align}
\bm{V}\bm{\mu} & =\bm{\mu}, \label{esag1} \\
|\bm{V}|& =1. \label{esag2}
\end{align}

As a result, it follows that the p.d.f. of the AG distribution presented in (\ref{esag_pdf}) can be expressed as:
\begin{align}
f_\text{ESAG}(\bm{y}) & =\frac{C_d}{B^{d/2}}
\exp \left [\frac{1}{2}\left \{\frac{\left (\bm{y}^\top \bm{\mu}\right)^2}{B}
-\bm{\mu}^\top \bm{\mu} \right \}\right ] \mathcal{M}_{d-1}\left \{\frac{\bm{y}^\top \bm{\mu}}
{B^{1/2}}  \right \} .
\label{esagden}
\end{align}

From (\ref{esag1}), it can be seen that the positive definite matrix $\bm{V}$ has a unit eigenvalue.  By defining the
other eigenvalues as
\begin{equation*}
0<\rho_1 \leq \cdots \leq  \rho_{d-1},
\end{equation*}
the inverse of  covariance matrix $\bm{V}$ can be written as:
\begin{equation*}
\bm{V}^{-1} = \xi_d\xi_d^\top + \sum_{j=1}^{d-1}  \xi_j \xi_j^\top/\rho_j,
\end{equation*}
where  $\xi_1$, \ldots , $\xi_{d-1}$ and $\xi_d=\mu/\|{\mu}\|$ is a set of mutually
orthogonal unit vectors and 
\begin{equation*}
\prod_{j=1}^{d-1} \rho_j =1.
\end{equation*}

For $d=3$, an efficient parameterization for the matrix $\bm{V}$ which satisfies the constraints (\ref{esag1}) and (\ref{esag2}) and has two free parameters. More specifically, let $\tilde\xi_1$ and $\tilde\xi_2$ be two  unit vectors orthogonal to each other and to the mean direction ${\xi}_3 = \bm{\mu} / \| \bm{\mu}\|$:
\begin{equation*}
\tilde\xi_1=\left (-\bm{\mu}_0^2, \mu_1 \mu_2, \mu_1 \mu_3\right )^\top/(\bm{\mu}_0\|\bm{\mu}\| ) \ \ \text{and} \ \ \tilde\xi_2= \left (0, -\mu_3, \mu_2\right)^\top/\mu_0,
\end{equation*}
where $\bm{\mu}=(\mu_1, \mu_2, \mu_3)^\top$ and $\mu_0=(\mu_2^2+\mu_3^2)^{1/2}$. Additionally,  the axes of symmetry $\xi_1$ and $\xi_2$ are expressed as:
\begin{equation*}
\xi_1=\cos \psi \, \tilde\xi_1 +  \sin \psi \,  \tilde\xi_2 \ \ \text{and} \ \ 
\xi_2=-\sin \psi \, \tilde\xi_1 + \cos \psi \,  \tilde\xi_2,
\end{equation*}
where $\psi \in (0, \pi]$ is the angle of rotation. Finally, by defining the pair ${\bm{\gamma}}=(\gamma_1, \gamma_2)^\top$ as
\begin{equation*}
\gamma_1=2^{-1}(\rho^{-1}-\rho)\cos 2 \psi \ \ \text{and} \ \
\gamma_2 =2^{-1}(\rho^{-1}-\rho) \sin 2 \psi,
\end{equation*}
where $\rho_1=\rho$ and $\rho_2=1/\rho$ for $\rho \in (0,1]$,  $\bm{V}^{-1}$ can be expressed as:
\begin{equation*}
\bm{V}^{-1} = \bm{I}_3+\gamma_1 \left (\tilde\xi_1 \tilde\xi_1^\top
-\tilde\xi_2 \tilde\xi_2^\top \right )
+\gamma_2 \left (\tilde\xi_1 \tilde\xi_2^\top
+\tilde\xi_2 \tilde\xi_1^\top \right ) + \left \{ (\gamma_1^2+\gamma_2^2+1)^{1/2}-1  \right \} \left (\tilde\xi_1 \tilde\xi_1^\top +\tilde\xi_2 \tilde\xi_2^\top \right ),
\end{equation*}
where $\bm{I}_3$ denotes the 3--dimensional identity matrix.

\subsection{The spherical elliptically symmetric projected Cauchy distribution}
\cite{tsagris2025b} projected the 3--dimensional Cauchy distribution on the sphere and, by imposing the same conditions as in ESAG (\ref{esag1})--(\ref{esag2}), introduced the spherical elliptically symmetric projected Cauchy distribution (SESPC) as:
\begin{eqnarray*} 
f_{\text{SESPC}}(\bm{y}) = \dfrac{B\left(\| \bm{\mu}\|^2+1\right)\sqrt{E}\left[\operatorname{arctan2}\left(\sqrt{E}, -\bf {y}^\top\bm{\mu}\right)-\operatorname{arctan2}\left(\sqrt{E}, \bf {y}^\top\bm{\mu}\right)+\pi\right]+2\bm{y}^\top\bm{\mu} E}{4\pi^2BE^2},
\end{eqnarray*}
where $E=B \|\bm{\mu}\|^2+B-(\bm{y}^\top\bm{\mu})^2$ and $\operatorname{arctan2}(.)$ is the 2--argument arc tangent. Similarly to the design of the ESAG distribution the corresponding inverse of the covariance matrix of $\bm{y}$, $\bm{V}$ is computed in the same exact manner.

\section{Model--based clustering} \label{sec:em}
Let $K$ be the number of clusters of a population where  $p_1,\ldots,p_K$ are the clusters'  unknown weights where $p_j>0$ and $\sum_{j=1}^{K}p_j=1$. Furthermore, we shall denote $\bm{Z}_i = (Z_{i1},\ldots,Z_{iK})^\top$, $i=1,\ldots,n$ as  the latent allocation variable following the  multinomial distribution,$\bm{Z}_i \sim\mathcal M(1;p_1,\ldots,p_K)$, with $K$ categories and success probabilities equal to $p_1,\ldots,p_K$ which are the corresponding  prior probabilities  of selecting an observation from cluster $j$. The distribution of $\bm{Y}_i$ conditionally on $\bm{Z}_i = \bm{z}_i$, can be expressed as:
\begin{equation}
\bm{Y}_i| (\bm{Z}_{i} = \bm{z}_i) \sim f_{\bm{Y}_i| \bm{Z}_{i}}(\bm{y}_i|\bm{z}_i) =\prod_{j=1}^{K} f^{z_{ij}}(\bm{y}_i;\bm{\mu}_j,\bm{\gamma}_j),\quad\mbox{independent for}\quad i = 1,\ldots,n
\end{equation}
where $f(\cdot;\bm{\mu}_j,\bm{\gamma}_j)$ denotes the probability density function of the  3-dimensional SESPC or ESAG distribution with  parameter $\bm{\mu}_j,\bm{\gamma}_j$ for $j=1,\ldots,K$. When $Z_{ij} = 1$, i.e.
 the $j$-th element of $\bm{Z}_{i}$ is equal to 1, the above equation can be written as  $\bm{Y}_i| (Z_{ij} = 1) \sim f(\cdot;\bm{\mu}_j)$. Additionally, the joint distribution of $\bm{Y}_i, \bm{Z}_i$ is: 
\begin{equation}
\label{eq:joint}
(\bm{Y}_i, \bm{Z}_{i}) \sim f_{\bm{Y}_i, \bm{Z}_i}(\bm{y}_i, \bm{z}_i) = \prod_{j=1}^{K}\left\{p_jf(\bm{y}_i;\bm{\mu}_j,\bm{\gamma}_j)\right\}^{z_{ij}}.
\end{equation}

while the marginal distribution of $\bm{Y}_i$ is expressed as  a mixture of $K$ distributions:
\begin{equation}
\label{eq:fmm}
\bm{Y}_i \sim f_{\bm{Y}_i}(\bm{y}_i) =\sum_{j=1}^{K}p_j f(\bm{y}_i;\bm{\mu}_j,\bm{\gamma}_j),\quad\mbox{independent for}\quad i = 1,\ldots,n.
\end{equation}

Furthermore, the conditional  probability that  an observation $i$ is assigned to component $j$, given that $\bm{Y}_i = \bm{y}_i$,  equals to 
\begin{equation}
\label{eq:post}
\mathrm{P}(Z_{ij} = 1|\bm{Y}_i = \bm{y}_i) =     \frac{p_j f(\bm{y}_i;\bm{\mu}_j,\bm{\gamma}_j)}{\sum_{\ell=1}^{K}p_\ell f(\bm{y}_i;\bm{\mu}_\ell,\bm{\gamma}_\ell)} =:w_{ij},\quad j=1,\ldots,K
\end{equation}
or equivalently, $\bm{Z}_i|(\bm{Y}_i = \bm{y}_i)\sim\mathcal M(1,w_{i1},\ldots,w_{iK})$, independent for $i=1,\ldots,n$. 

Finally, the  observed log--likelihood is defined as 
\begin{equation}
\label{eq:mixlik}
\ell(\bm{\mu},\bm{\gamma}, \bm{p}|\bm{y}) = \sum_{i=1}^{n}\log\left\{\sum_{j=1}^{K}p_j f(\bm{y}_i;\bm{\mu}_j,\bm{\gamma}_j)\right\},
\end{equation}
where $\bm{\mu},\bm{\gamma} \in\mathbb R^{4K}$, $\bm{p}\in\mathcal P_{K-1}$, where $\mathcal P_{K-1} = \{p_1,\ldots,p_K: p_j > 0, j=1,\ldots,K, \sum_{j=1}^{K}p_j = 1\}$. It is clear that the goal is to maximize \eqref{eq:mixlik}. In order to achieve that, similarly to \cite{tsagris2025a}, the EM algorithm is considered where for a set of starting values,  the expectation of the complete log--likelihood (E--step) is computed and then is maximized with respect to $\bm{\mu}$, $\bm{p}$ and $\bm{\gamma}$ (M--step). In particular, the complete log--likelihood is
\begin{equation}
\ell_c(\mu,\, \gamma,\, p \mid y, z) = \sum_{i=1}^{n} \sum_{j=1}^{K}z_{ij}\left[
\log p_j + \log f(y_i;\, \mu_j,\, \gamma_j) \right].
\label{eq:complete_loglik}
\end{equation}

\noindent\textbf{E--step.}
At iteration $t$, given the current parameter estimates $\Theta^{(t)} = \{\mu^{(t)},\, \gamma^{(t)},\, p^{(t)}\}$, the expected value of $Z_{ij}$  is computed with respect to the conditional distribution $Z_i \mid (Y_i = y_i,\, \Theta^{(t)})$. The corresponding  membership probabilities are derived as:
\begin{equation}
w_{ij}^{(t)}=\frac{p_j^{(t)}\, f\!\left(y_i;\, \mu_j^{(t)},\, \gamma_j^{(t)}\right)}
{\displaystyle\sum_{\ell=1}^{K} p_\ell^{(t)}\, f\!\left(y_i;\, \mu_\ell^{(t)},\, \gamma_\ell^{(t)}\right)}, \qquad i = 1,\ldots,n,\quad j = 1,\ldots,K.
\label{eq:Estep}
\end{equation}
 
\medskip
\noindent\textbf{M--step.}
At this step,  the expected complete log-likelihood
$Q(\Theta \mid \Theta^{(t)}) = \mathbb{E}[\ell_c \mid y,\, \Theta^{(t)}]$ is maximised with respect to $\Theta$ and the mixing proportions are updated  as the weighted  averages of the posterior probabilities:
\begin{equation}
p_j^{(t+1)}=\frac{1}{n} \sum_{i=1}^{n} w_{ij}^{(t)}, \qquad j = 1, \ldots, K.
\label{eq:update_pj}
\end{equation}
Due to the fact that neither the ESAG nor the SESPC density yields a closed--form maximizer for $(\mu_j,\, \gamma_j)$, the weighted log--likelihood contribution of component $j$,
\begin{equation}
Q_j(\mu_j,\, \gamma_j)= \sum_{i=1}^{n}w_{ij}^{(t)} \log f(y_i;\, \mu_j,\, \gamma_j),
\label{eq:Qj}
\end{equation}
is maximized numerically. Specifically, for each component $j = 1, \ldots, K$, a general--purpose
optimizer (e.g. Nelder--Mead) is applied to $Q_j$, and the procedure is restarted until the improvement in the objective falls below $\epsilon$ ($=10^{-5}$), yielding the updated estimates $\mu_j^{(t+1)}$ and $\gamma_j^{(t+1)}$.

Regarding the optimal number of clusters the integrated complete likelihood (ICL) has been considered (\cite{biernacki2003choosing}. Initialization of the EM algorithm demands extra care  in order to avoid convergence to local maxima. Some indicative works on this topic include \cite{biernacki2003choosing, karlis2003choosing, fraley2005incremental, baudry2015mixtures, papastamoulis2016estimation, michael2016effective}. In our implementation of the mixtures of ESAG and SESPC distributions, we tried two initialization schemes. One based on $k$--means starting values and one based on Gaussian mixture models (GMMs). The first strategy is adequate for rotationally symmetric distributions \citep{tsagris2025a}, but in our case GMMs that allow for a structured covariance was shown to yield better performance. 

\subsection{Inclusion of covariates}
We may link the mixing probabilities $p_j$ to some covariates $\bm{X}$ yielding the concomitant model
\begin{equation}
\label{eq:post2}
p_j(\bm{x}_i) \;=\; \frac{\exp{\left(\bm{x}^\top_i\bm{\beta}_j\right)}}{\sum_{\ell=1}^K\exp{\left(\bm{x}^\top_i\bm{\beta}_\ell\right)}}.
\end{equation}
In this case we substitute the $p_j$ in Eq (\ref{eq:update_pj}) with $p_j(\bm{x}_i)$ from Eq. (\ref{eq:post2}), and hence the M--step requires an extra optimization. The $\bm{\beta}_{js}$ are obtained from maximization of the following quantity 
\begin{equation*}
D_j \;=\; \sum_{i=1}^{n_j}w^{(t)}_{ij}\log\left[p_j\left(\bm{x}_i\right)\right], \qquad j = 1, \ldots, K. 
\end{equation*}
The maximization of $D_j$ is performed independently of the maximization of the weighted log--likelihood in Eq. (\ref{eq:Qj}), and the observed log--likelihood (\ref{eq:mixlik}) now becomes
\begin{equation*}
\ell(\bm{\mu},\bm{\gamma}, \bm{p}|\bm{y}) = \sum_{i=1}^{n}\log\left\{\sum_{j=1}^{K}p_j\left(\bm{x}_i\right) f(\bm{y}_i;\bm{\mu}_j,\bm{\gamma}_j)\right\}.
\end{equation*}

We may also link the mean vector of the ESAG or SESPC to some covariates yielding the mixtures of regressions model, where $\bm{\mu}_j(\bm{x}_i)=\bm{x}_i^\top\bm{\beta}_j$. In this case, the observed log--likelihood becomes
\begin{equation*}
\ell(\bm{\mu},\bm{\gamma}, \bm{p}|\bm{y}) = \sum_{i=1}^{n}\log\left\{\sum_{j=1}^{K}p_jf(\bm{y}_i;\bm{\mu}_j\left(\bm{x}_i^\top\bm{\beta}_j\right),\bm{\gamma}_j)\right\}.
\end{equation*}
Both the ESAG and the SESPC offer the possibility to relate their $\gamma$ parameters to some covariates in the same linear manner, at the cost of increased computational cost. The last option is to add the covariates in both the mixing probabilities and the parameters of either distribution, but this is computationally expensive. 

\subsection{Visual inspection of the mixtures of ESAG and SESPC distributions}
Our scope is to investigate the performance of each model  assuming two scenarios where the data comes from the ESAG distribution and  the data comes from the SESPC distribution. For each case, under the assumption that the true number of clusters is two, we will compare the performance of each model using simulation based on their corresponding ARI value. Before we proceed to any numerical investigations, it is important to examine the effect on the vector parameter $\gamma$ on forming a cluster. Regarding the structure of the clusters, the corresponding mean vectors were defined as follows. For a given value of $\bm{\mu_1}$, by expressing it through its corresponding unit vector $\bm{\mu_1}'=\frac{\bm{\mu_1}}{||\bm{\mu_1}||}$, we find a  vector $\bm{\mu_2}'$ such that the angle between $\bm{\mu_1}'$ and $\bm{\mu_2}'$ is equal to $\omega$. Lastly, we simulate a dataset forming two clusters by generating a random sample of size $n=200$ from a mixture of ESAG (SESPC) distributions with approximately equal mixing probabilities, $p_1 \approx p_2$. For each cluster the corresponding mean vectors are considered to be equal to $\tau_1\times \bm{\mu_1}'$ and $\tau_2\times \bm{\mu_2}'$ respectively. As it will be shown hereafter, the values of $(\tau_1,\tau_2)$ play the role of the concentration parameter.

\begin{figure}
\begin{tabular}{ccc}
\multicolumn{3}{c}{\underline{ESAG distribution}}  \\
\includegraphics[scale=0.4, trim = 40 0 0 0]{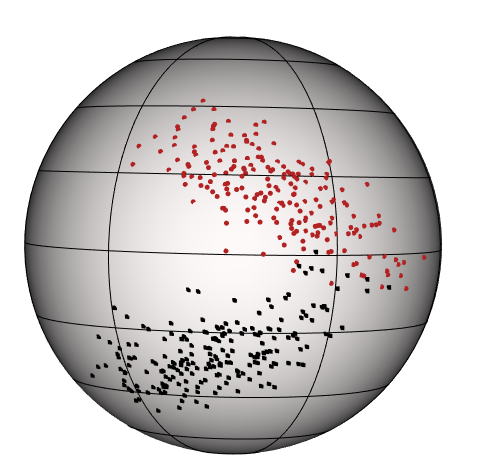} &  %
\includegraphics[scale=0.4, trim = 40 0 0 0]{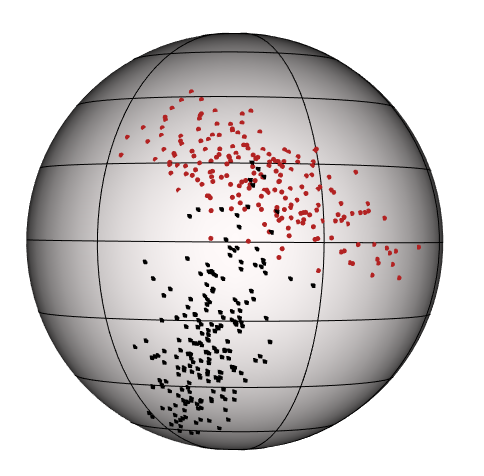} &
\includegraphics[scale=0.4, trim = 40 0 0 0]{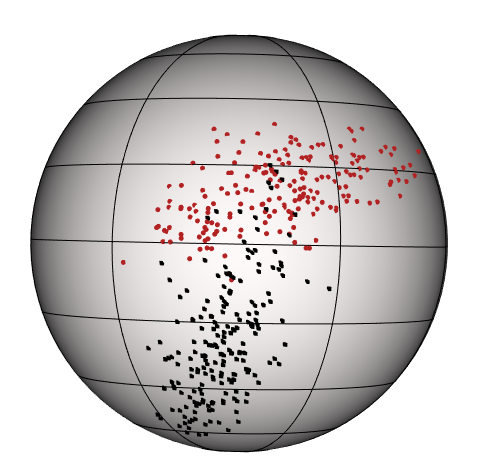} \\
$\bm{\gamma}=[(1,1),(1,1)]$ & $\bm{\gamma}=[(1,1)^\top,(-1,1)^\top]$ & $\bm{\gamma}=[(-1,1)^\top,(-1, 1)^\top]$ \\
\multicolumn{3}{c}{\underline{SESPC distribution}}  \\
\includegraphics[scale=0.4, trim = 40 0 0 0]{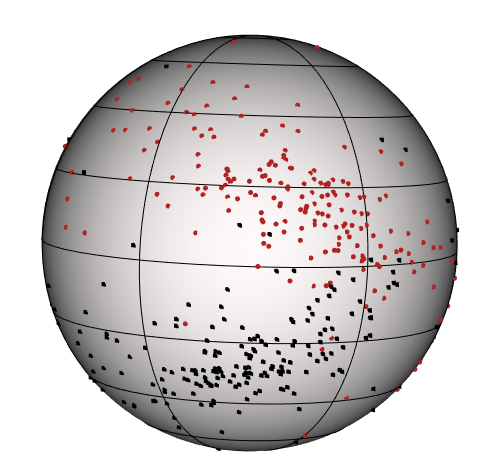} & 
\includegraphics[scale=0.4, trim = 40 0 0 0]{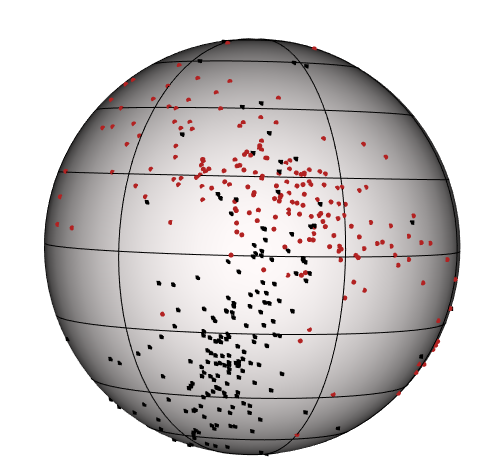} &
\includegraphics[scale=0.4, trim = 40 0 0 0]{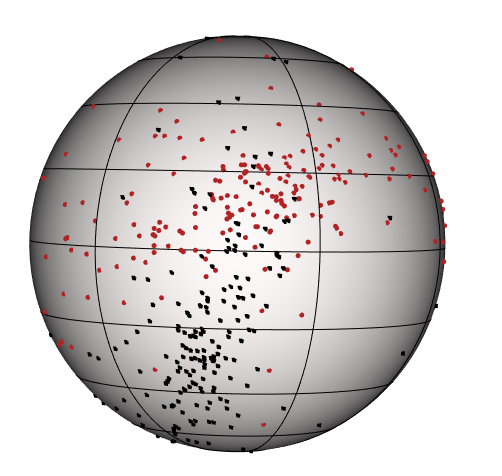} \\
$\bm{\gamma}=[(1,1)^\top,(1,1)^\top]$ & $\bm{\gamma}=[(1,1)^\top,(-1, 1)^\top]$ & $\bm{\gamma}=[(-1,1)^\top,(-1,1)^\top]$
\end{tabular}
\caption{Effect of $\bm{\gamma}$ for ESAG distribution (top row, with $(\tau_1,\tau_2)=(5,5)$) and SESPC  distribution (bottom row, with $(\tau_1,\tau_2)=(5,5)$). For all cases, $\bm{\mu_1'}=(-0.927,-0.282,0.249)^\top$, $\bm{\mu_2'}=(-0.886,-0.008,-0.464)^\top$.}
\label{effect_of_gamma}
\end{figure}

In Figure \ref{effect_of_gamma} the corresponding shape of two clusters is presented under different values of $\bm{\gamma}$. For all cases, we set $\omega=45$, $\bm{\mu_1'}=(-0.927,-0.282,0.249)$ and $\bm{\mu_2'}=(-0.886, -0.008,-0.464)$, by generating $\bm{\mu_1}$ based on three random numbers from  $N(0,1)$. For illustration purposes we fix the values of $(\tau_1,\tau_2)$ at $(5,5)$. From Figure \ref{effect_of_gamma}, it can be seen that the values of $\gamma$ affect how the two clusters are formed. Note that, we intentionally considered these specific values of $\gamma$ in order to investigate each model's performance under the scenarios where the two clusters are formed vertically or in an angle. Additionally, it can be observed that, considering exactly the same parameters for both distribution, the SESPC distribution is more spread than the ESAG but the corresponding relation between the two clusters remains the same.
 
\subsection{Model--based clustering with hyper--spherical data}
From Section \ref{sec:general}, it is clear that the parameterization of ESAG and SESPC distributions, as they were originally introduced in the works of \cite{paine2018} and \cite{tsagris2025b}, can be considered only for spherical data. As a consequence, an interesting research question arises which is how practitioners could consider the use of Model--based clustering for hyper--spherical  elliptical data. Recently, \cite{yu2024} proposed a novel parameterization of the ESAG distribution for hyper--spherical data. Motivated by their work, we consider the EM algorithm presented in Section \ref{sec:em}. However, some preliminary results showed that  the computational cost of the EM is quite  high. 

In order to tackle this limitation, we suggest to project the data onto the sphere using a geodesic principal component analysis (PCA) type of projection \citep{fletcher2004,jung2012} and then apply the SESPC or ESAG mixture model. The algorithm first computes the intrinsic Fréchet mean of the data points on the sphere, then uses the logarithmic map to project all points into the tangent space at that mean. In this flat tangent space, standard PCA is applied to find the top principal directions. The function then constructs a low--dimensional geodesic subspace (a 3--dimensional sphere) by orthonormalizing the mean vector with the principal tangent directions, and finally projects the original spherical data onto this subspace, yielding points on the sphere. 

\begin{figure}
\begin{tabular}{ccc}
\includegraphics[scale=0.32, trim = 40 0 0 0]{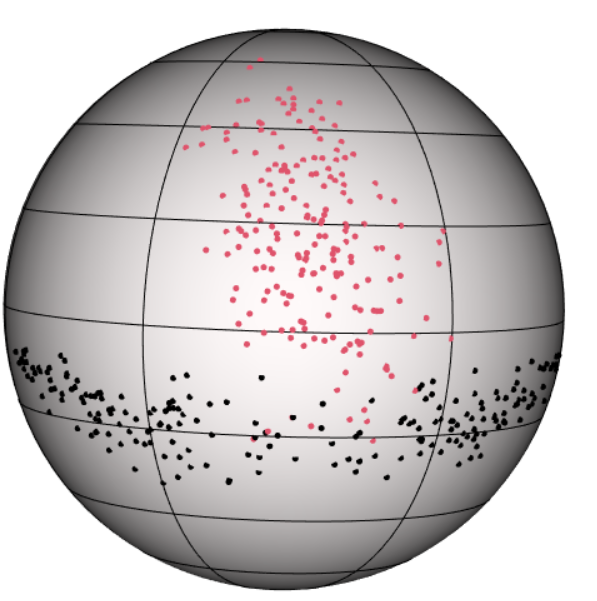} &  %
\includegraphics[scale=0.32, trim = 40 0 0 0]{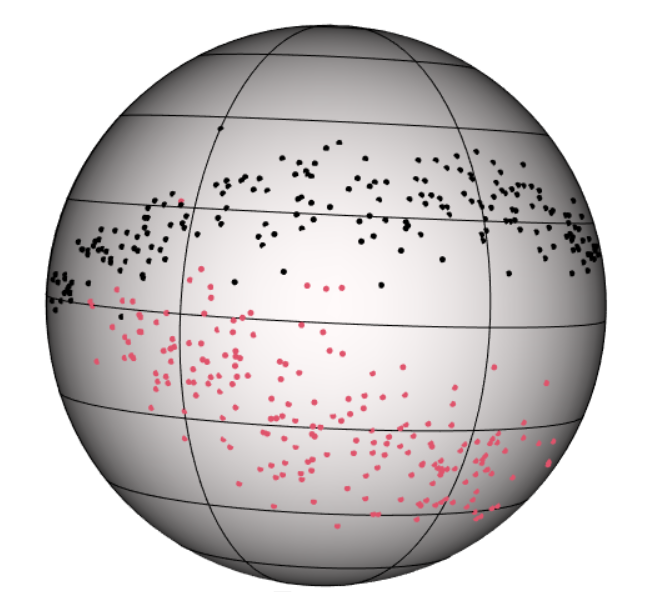} &
\includegraphics[scale=0.32, trim = 40 0 0 0]{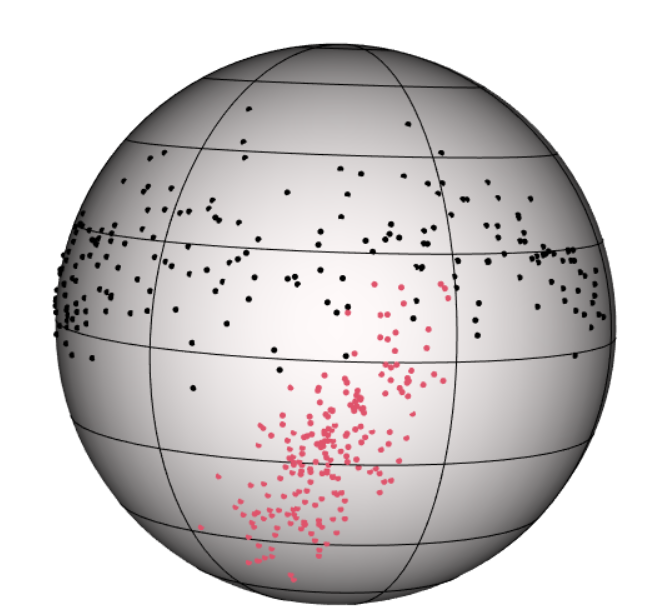} \\
Case 1 & Case 2 & Case 3 \\
\end{tabular}
\caption{Hyper--spherical data for the projected 10--dimensional  ESAG distribution. The values of $\bm{\gamma_1},\bm{\gamma_2}$ and $\bm{\mu_1},\bm{\mu_2} $ are presented in Figures \ref{values_mu_hyper} and \ref{values_gam_hyper}, respectively, in the Appendix.}
\label{hyper_effect_of_gamma}
\end{figure}

In Figure \ref{hyper_effect_of_gamma} several sphere plots are illustrated of the projected 10-dimensional data into the sphere. The corresponding parameter $\bm{\gamma_1}$ , $\bm{\gamma_2}$ and $\bm{\mu_1}$, $\bm{\mu_2} $ are presented in Figures \ref{values_mu_hyper} and \ref{values_gam_hyper}, respectively, in the Appendix.

\section{Simulation studies} \label{sec:simulations}
\subsection{Numerical analysis for spherical data}
\begin{figure}
\centering
\includegraphics[width=0.46\textwidth]{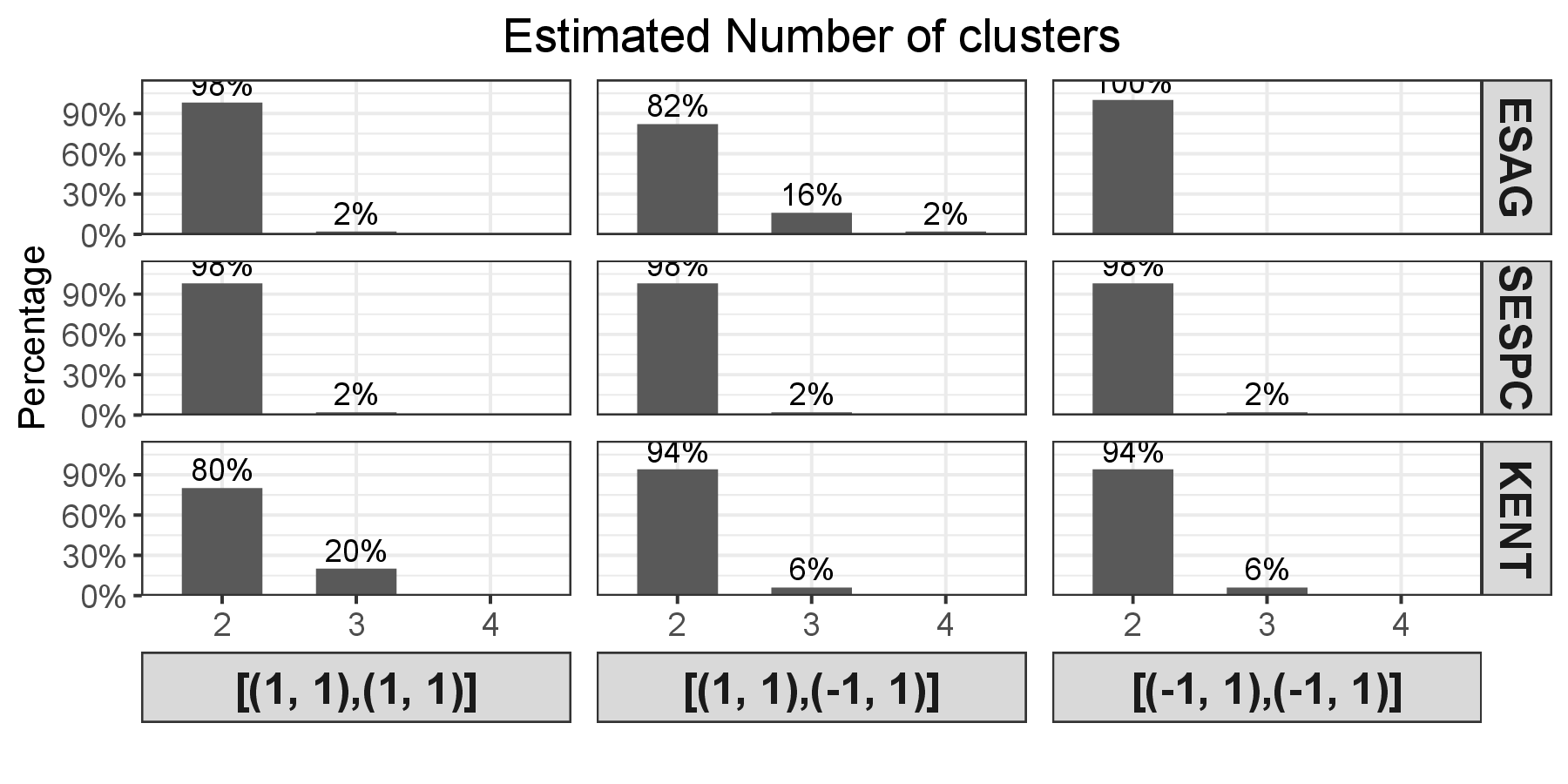} 
\includegraphics[width=0.5\textwidth]{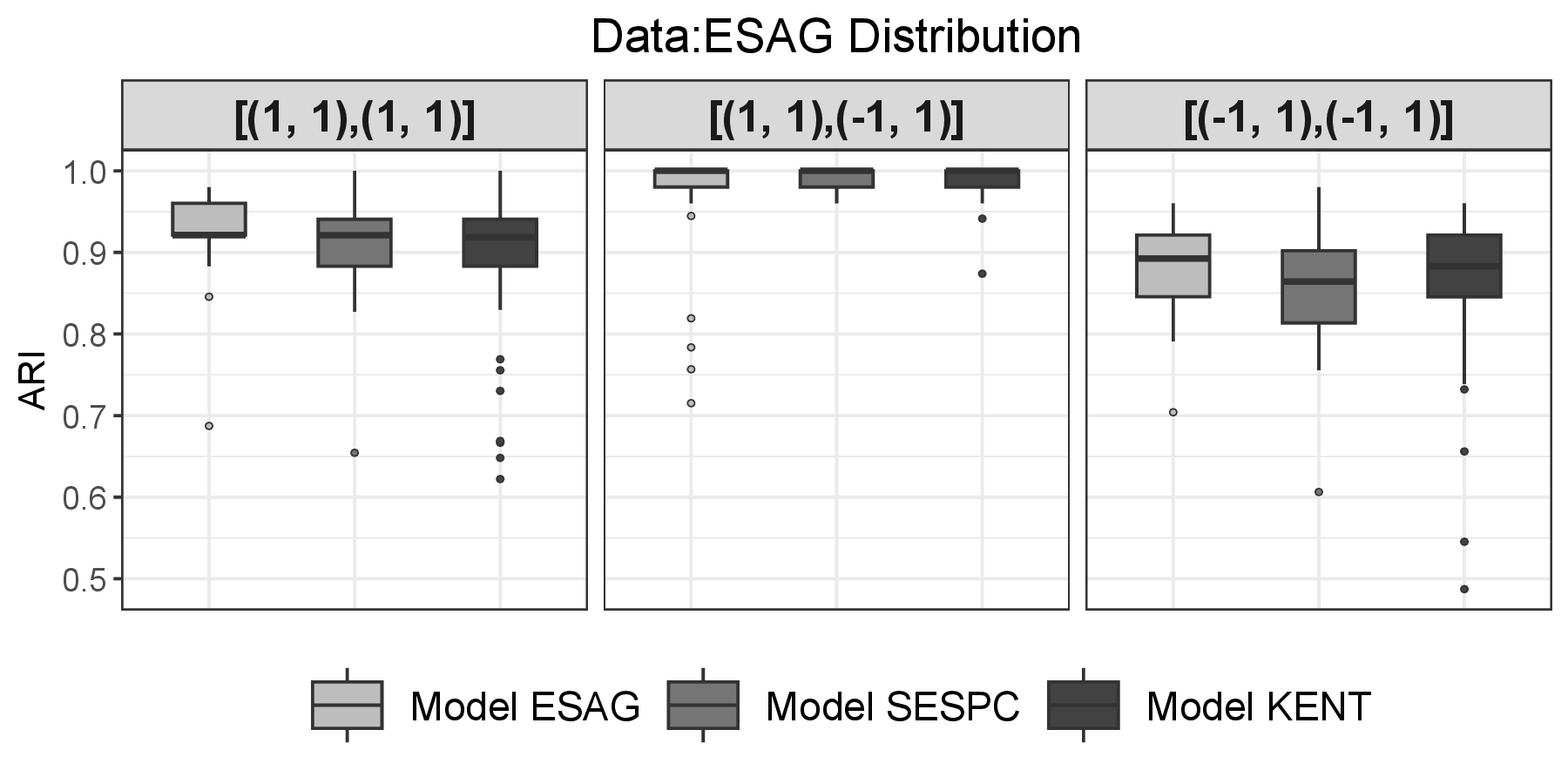} \\
\includegraphics[width=0.46\textwidth]{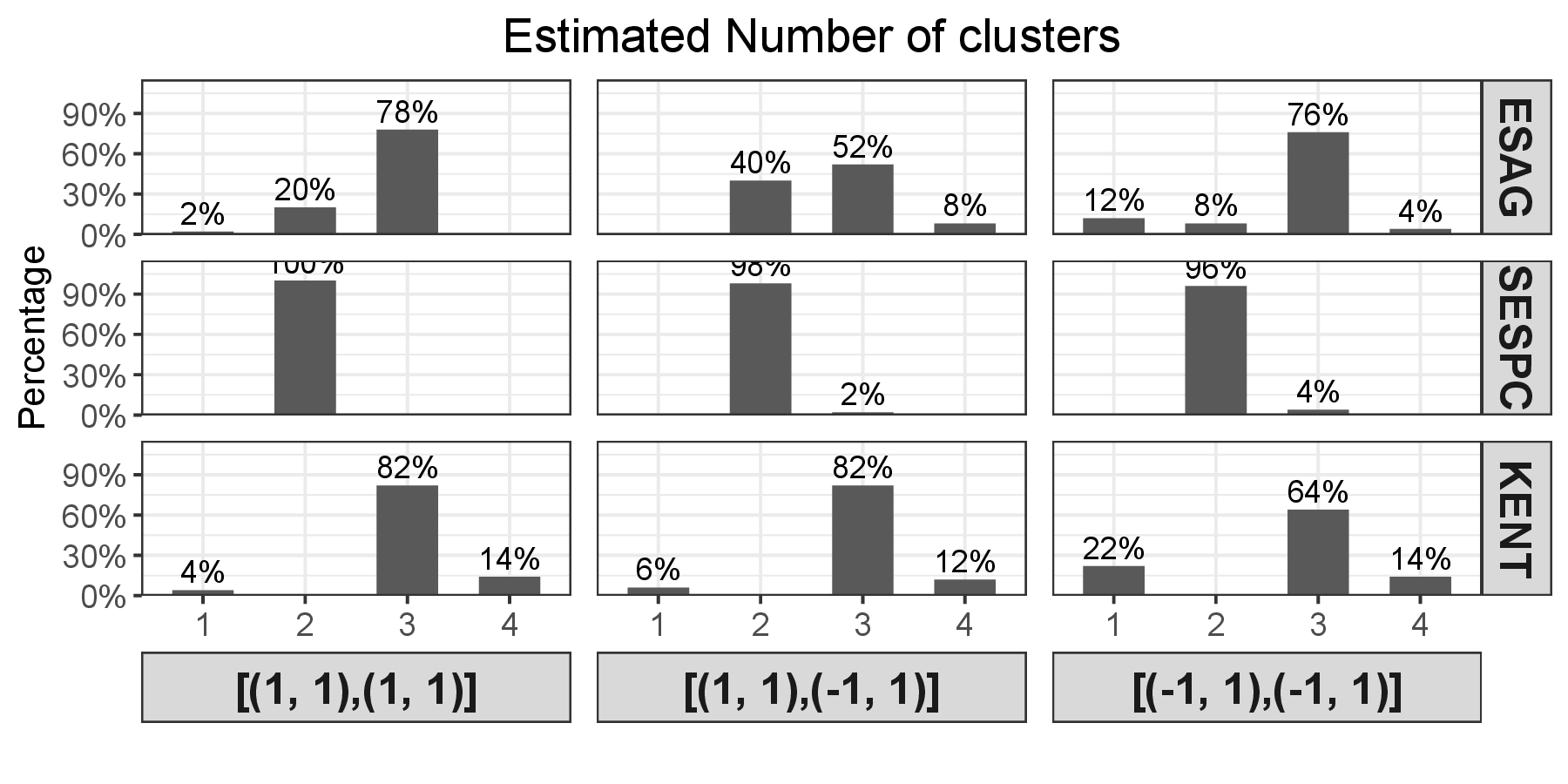} 
\includegraphics[width=0.5\textwidth]{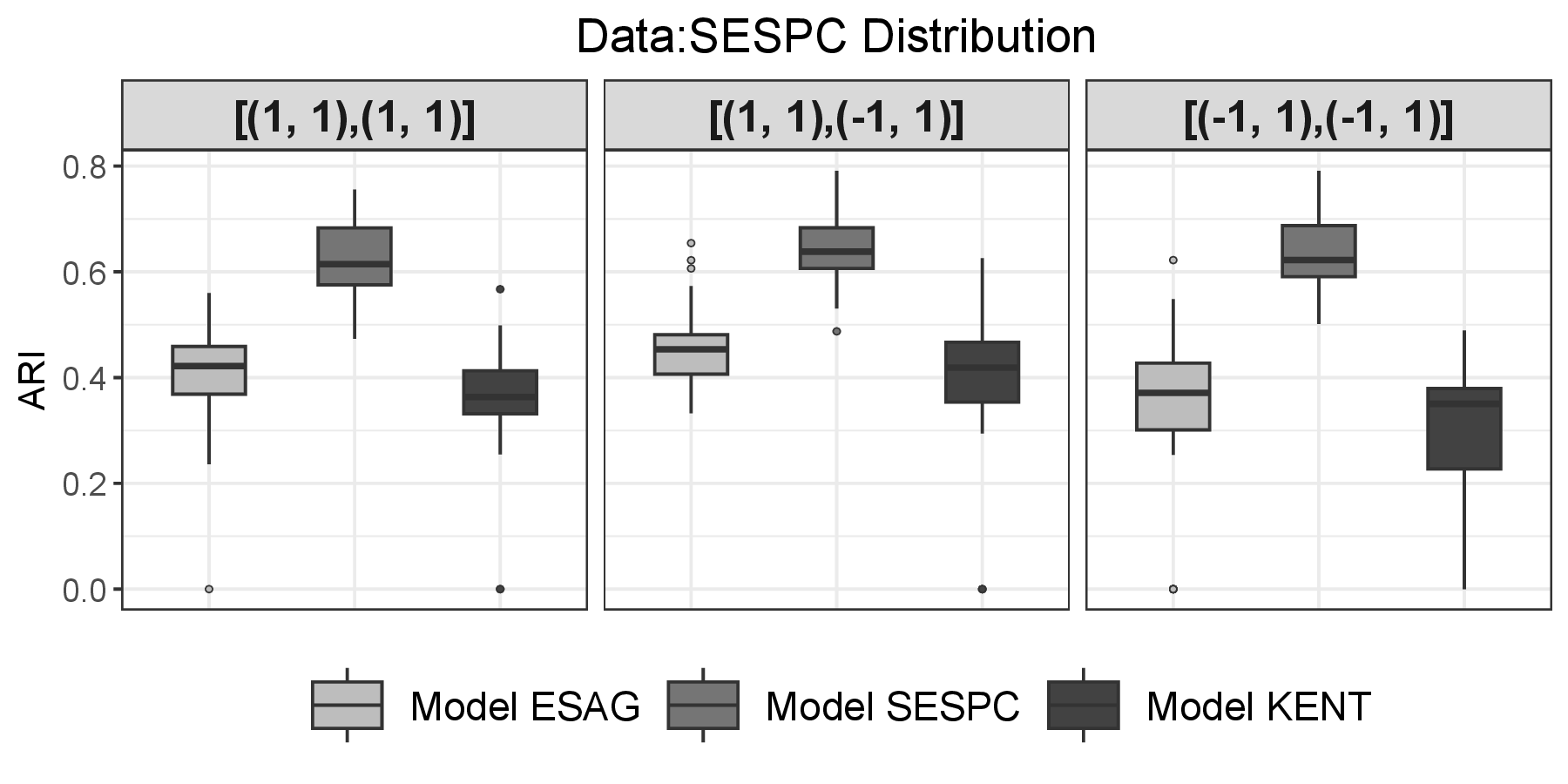} 
\caption{ESAG and SESPC generated data with equal clusters: $n=200$, $\omega=45^\circ$, $\bm{\mu_1'}=(-0.927,-0.282,0.249)^\top$, $\bm{\mu_2'}=(-0.886, -0.008,-0.464)^\top$, $(\tau_1,\tau_2) = (5,5)$, 
and various $\gamma$ combinations.}
\label{n200_equal_prob_D_ESPC}
\end{figure}

Firstly, the performance of each method, assuming  almost equal sizes for the two clusters, is investigated (total size $n=200$). In Figure \ref{n200_equal_prob_D_ESPC}, the performance  of the two models are presented. In particular, for $R=50$ iterations (i.e. simulating 50 different samples consisting of two clusters) the boxplots of the Adjusted Rand index (ARI) metric are presented. For simplicity, in the results presented below we shall denote as $ARI^{ESAG}$ ($ARI^{SESPC}$) the ARI values obtained considering the ESAG (SESPC) model. Furthermore, In the performance comparisons presented in the remainder of this Section, we shall consider also  the use of the mixtures of Kent distributions \citep{peel2001} as an additional competitor. Note that, based on the existing literature \citep{paine2018} we expect to see that ESAG and Kent mixture models will have similar performance.  We may see that:
 
\begin{itemize}
\item ESAG distribution: When the model being considered is the ESAG, it can create the correct number of clusters in  all the cases  where large values of ARI indicate the correct number of predicted clusters. Additionally, when the model is the SESPC, the performance does not significantly change. In particular, for the case  where $\bm{\gamma}=[(1,1)^\top,(1,1)^\top]$ ($\bm{\gamma}=[(1,1)^\top,(-1,1)^\top]$ ), the boxplots of $ARI^{ESAG}$'s are narrowed comparing to the boxplots of $ARI^{SESPC}$ with corresponding medians close to 0.95 (1). Lastly, when $\bm{\gamma}=[(-1,1)^\top,(-1,1)^\top]$ the  boxplots of $ARI^{ESAG}$'s are wider comparing to the boxplots of $ARI^{SESPC}$ with corresponding medians close to 0.9. In summary, for all the investigated scenarios, both models  can detect the correct number of clusters and give similar  ARI values larger than 0.8 while the  worst performance is achieved  under the third case.  
 
\item SESPC distribution: Using the SESPC  model can effectively define the  correct number of clusters in most of the cases. Furthermore, for all the investigated cases, the boxplots of $ARI^{SESPC}$ do not significantly differ. On the other hand, using the ESAG model the efficiency on detecting the correct number of clusters is worse and the boxplots of $ARI^{ESAG}$ indicate poor performance with large variability in the last case. 

\item Performance comparisons: The performance of each model is better when the data comes from the ESAG distribution, which is an expected result since in both cases the same values of $\bm{\tau}=(5,5)$ are considered. Let us recall the fact that the SESPC  distribution is more spread in the tails compared to the ESAG (see Figure \ref{effect_of_gamma}). As expected, regardless the underlying distribution of the data, ESAG and Kent models have almost the same performance.

\end{itemize}

\begin{figure}
\centering
\includegraphics[width=0.46\textwidth]{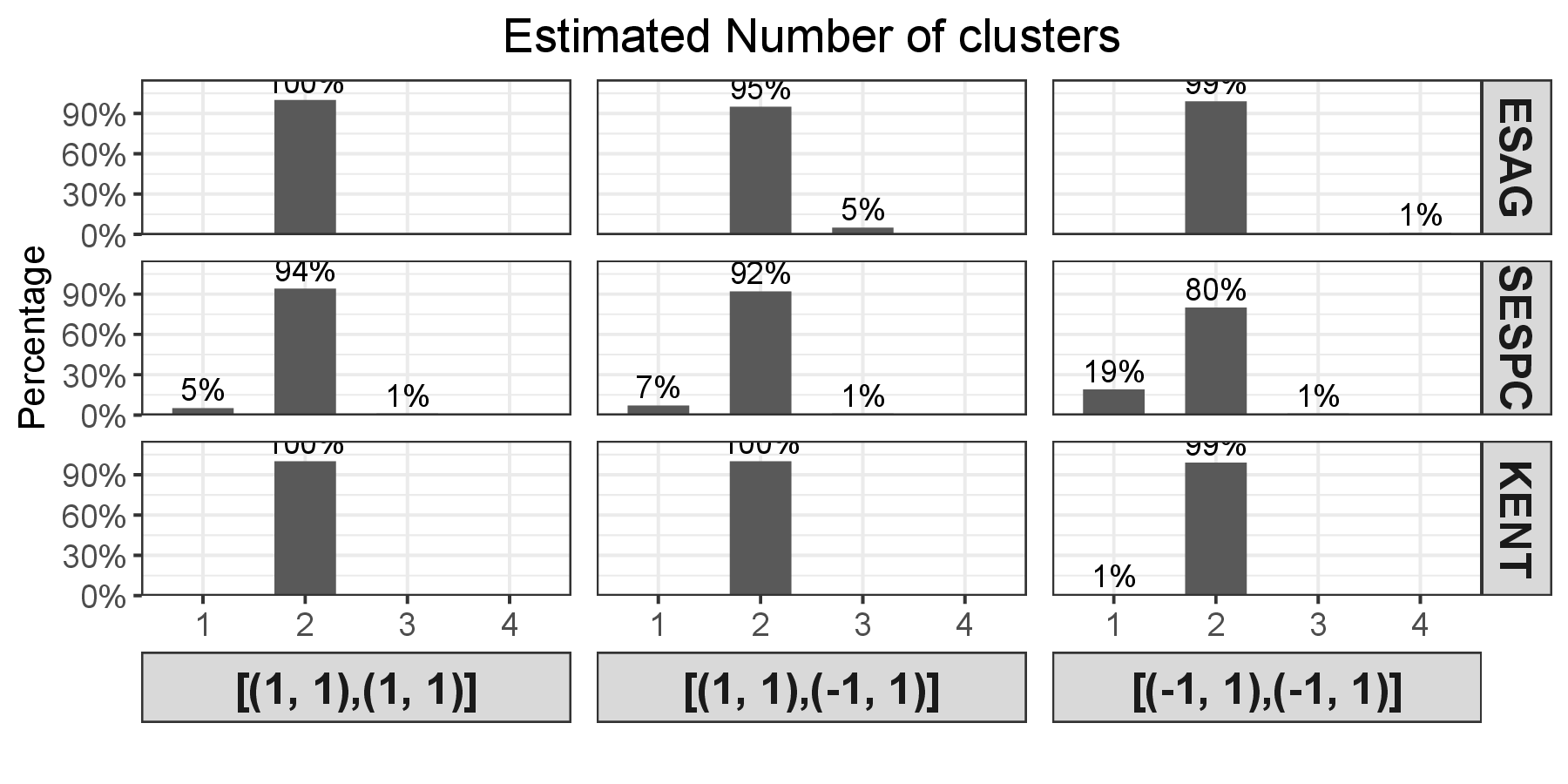} 
\includegraphics[width=0.53\textwidth]{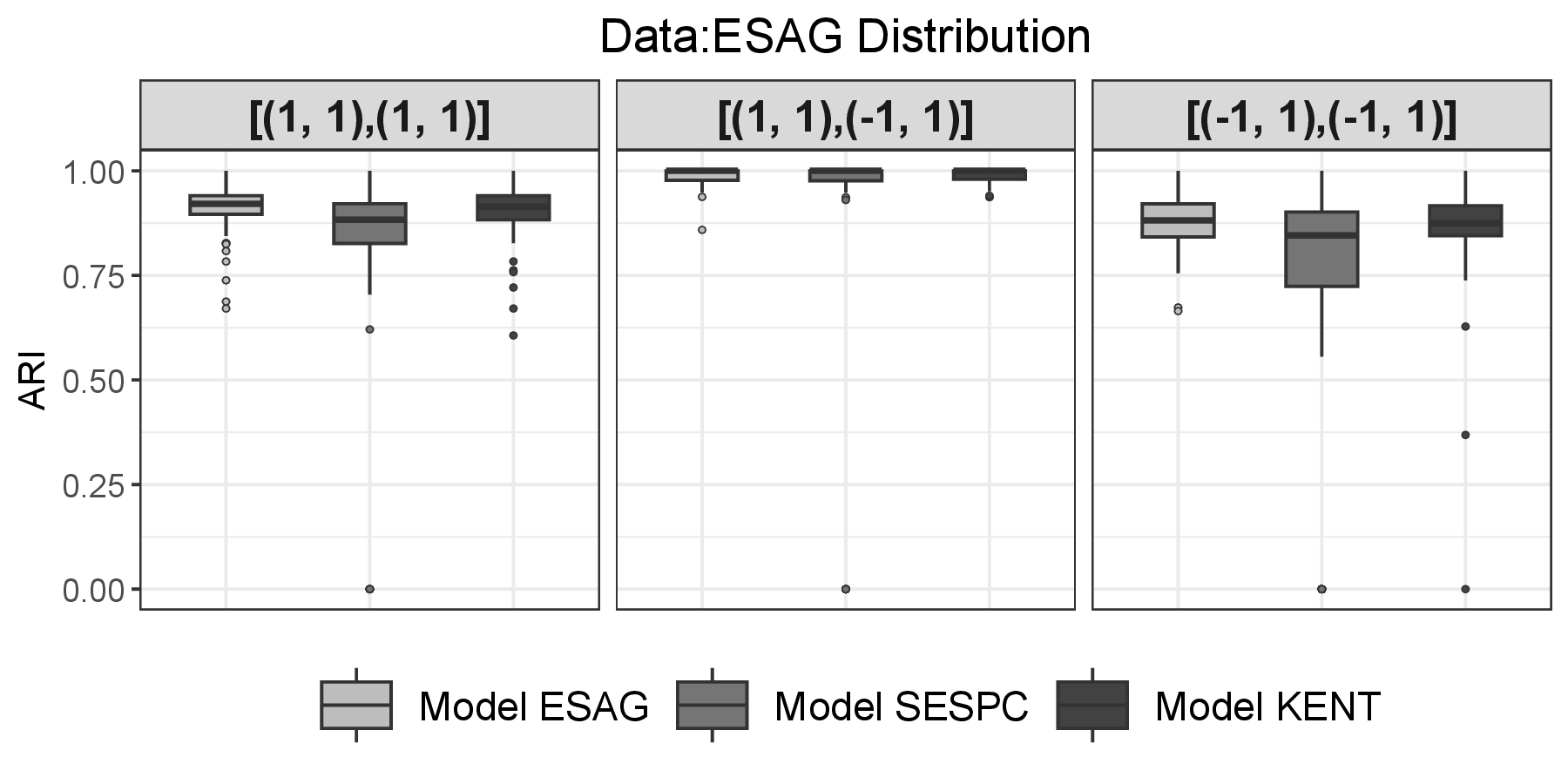} 
\includegraphics[width=0.46\textwidth]{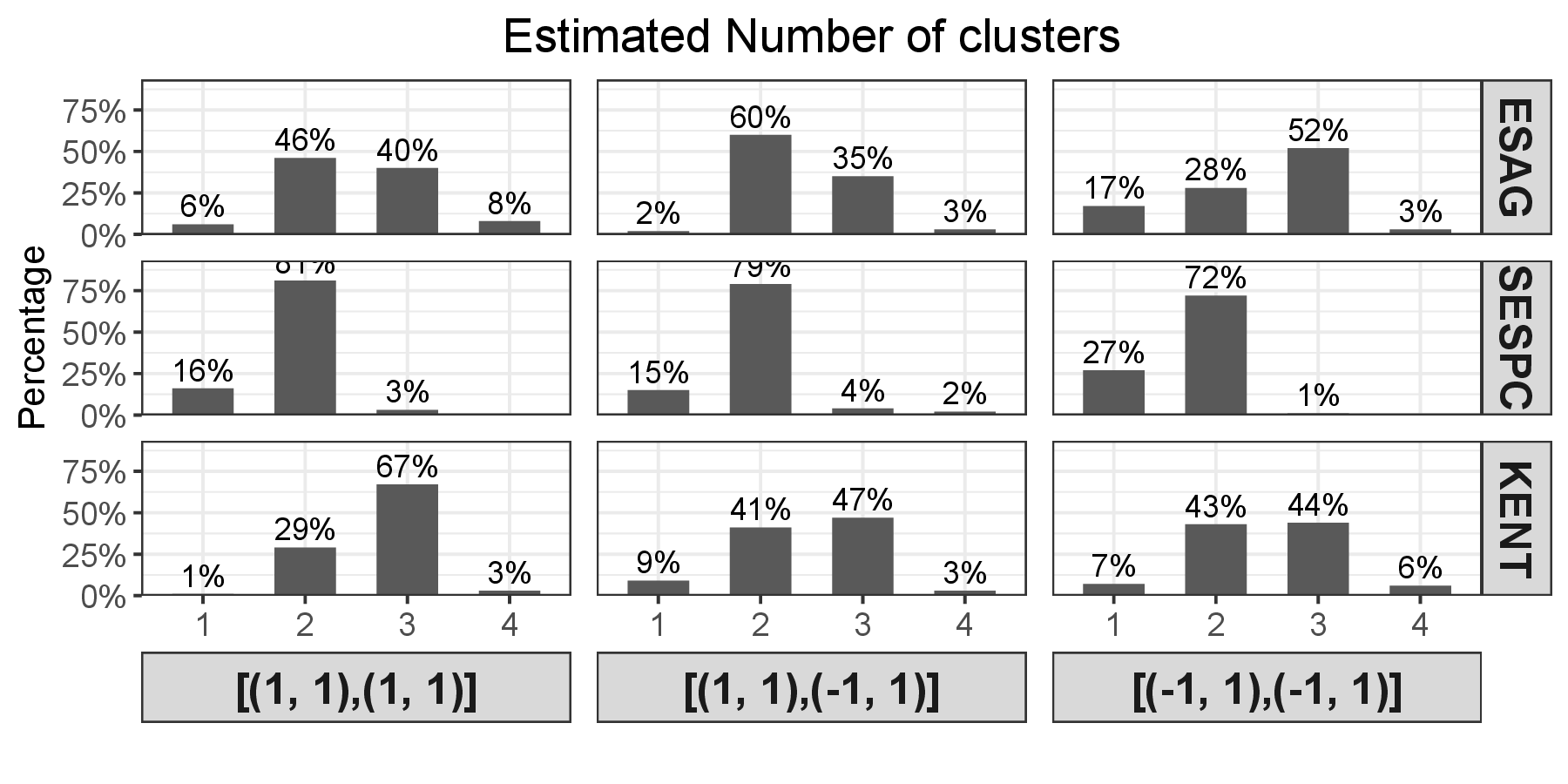} 
\includegraphics[width=0.53\textwidth]{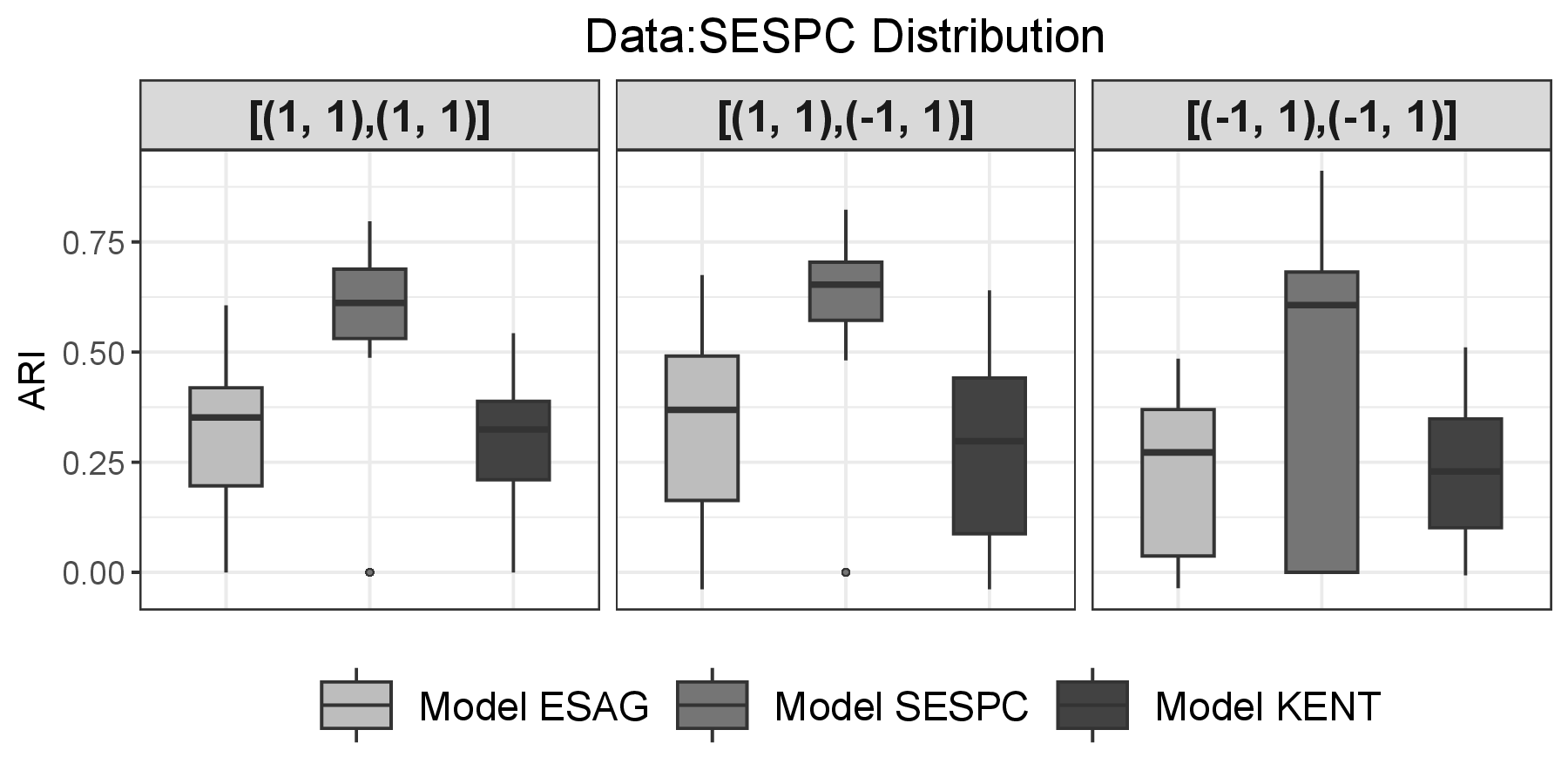} 
\caption{ESAG and SESPC generated data with unequal clusters: $n=200$,  $\omega=45^\circ$, $\bm{\mu_1'}=( -0.927,-0.282,0.249)^\top$, $\bm{\mu_2'}=(-0.886, -0.008,-0.464)^\top$, $(\tau_1,\tau_2)=(5,5)$, and various $\gamma$ combinations.}
\label{n200_unequal_prob_D_ESAG}
\end{figure}

In Figure \ref{n200_unequal_prob_D_ESAG}, some additional performance comparisons between the two models are illustrated but now the probabilities of an observation to belong in a cluster is randomly generated from the 2--dimensional Dirichlet distribution. Note that, we intentionally increase the number of iterations on $R=100$ in order to capture many scenarios with respect to the size of each cluster. From  Figure \ref{n200_unequal_prob_D_ESAG} we observe that:

\begin{itemize}
\item ESAG distribution: When the model being considered is the ESAG,  it can create the correct number of clusters in  almost all the cases and the corresponding boxplots of $ARI^{ESAG}$ do not significantly differ. On the other hand, when the model is the SESPC, the performance does not significantly change for the first two cases while for the third one, the variability  of $ARI^{SESPC}$ is very large. 
 
\item SESPC distribution: The SESPC  model can effectively  define the  correct number of clusters in most of the cases. The smallest variability in the boxplots $ARI^{SESPC}$ is achieved for the case where $\bm{\gamma}=[(1,1)^\top,(1,1)^\top]$ while for the other two the variability is quite large. It seems that large difference between the size of the clusters  might affect the performance of the SESPC model in some cases. Nevertheless, it should be noted that, using the ESAG model, the efficiency on detecting the correct number of clusters is  significantly low and the boxplots of $ARI^{ESAG}$ indicate worse performance comparing to the SESPC model.

\item Performance comparisons: The performance of each model is better when the data comes from the ESAG distribution, which is an expected result since in both cases the same values of $\bm{\tau}=(5,5)$ are considered and the SESPC  is more spread in the tails compared to the ESAG distribution (see, Figure \ref{effect_of_gamma}). Similarly to the case of having equal clusters, we observe that ESAG and Kent models have almost the same performance. In summary, under the assumption that the sizes of the clusters are balanced, the performance between the two models is more or less  the same with the difference that when the data come from the SESPC distribution the ESAG model performs worse.
\end{itemize}

\begin{figure}
\centering
\includegraphics[scale=0.3]{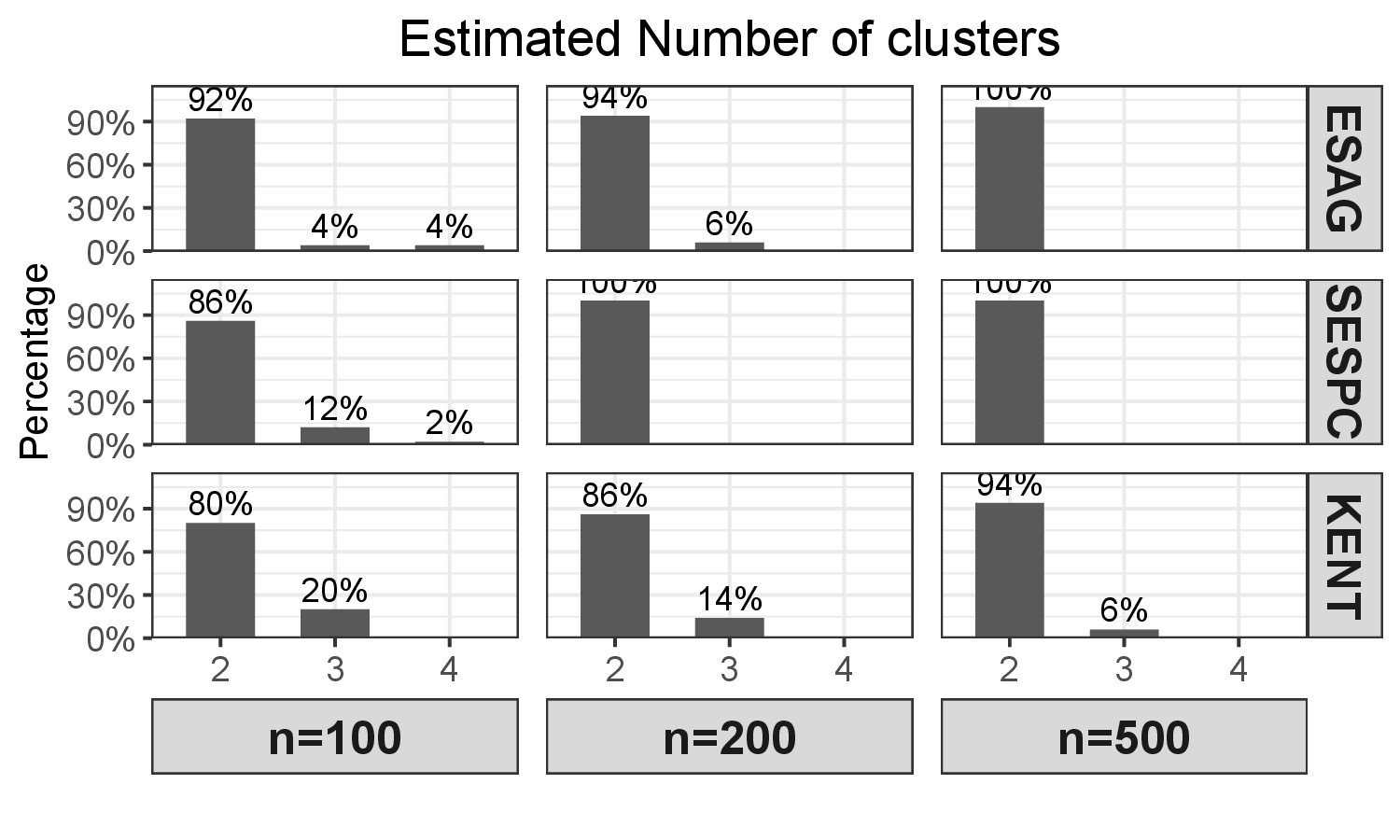}   \includegraphics[scale=0.3]{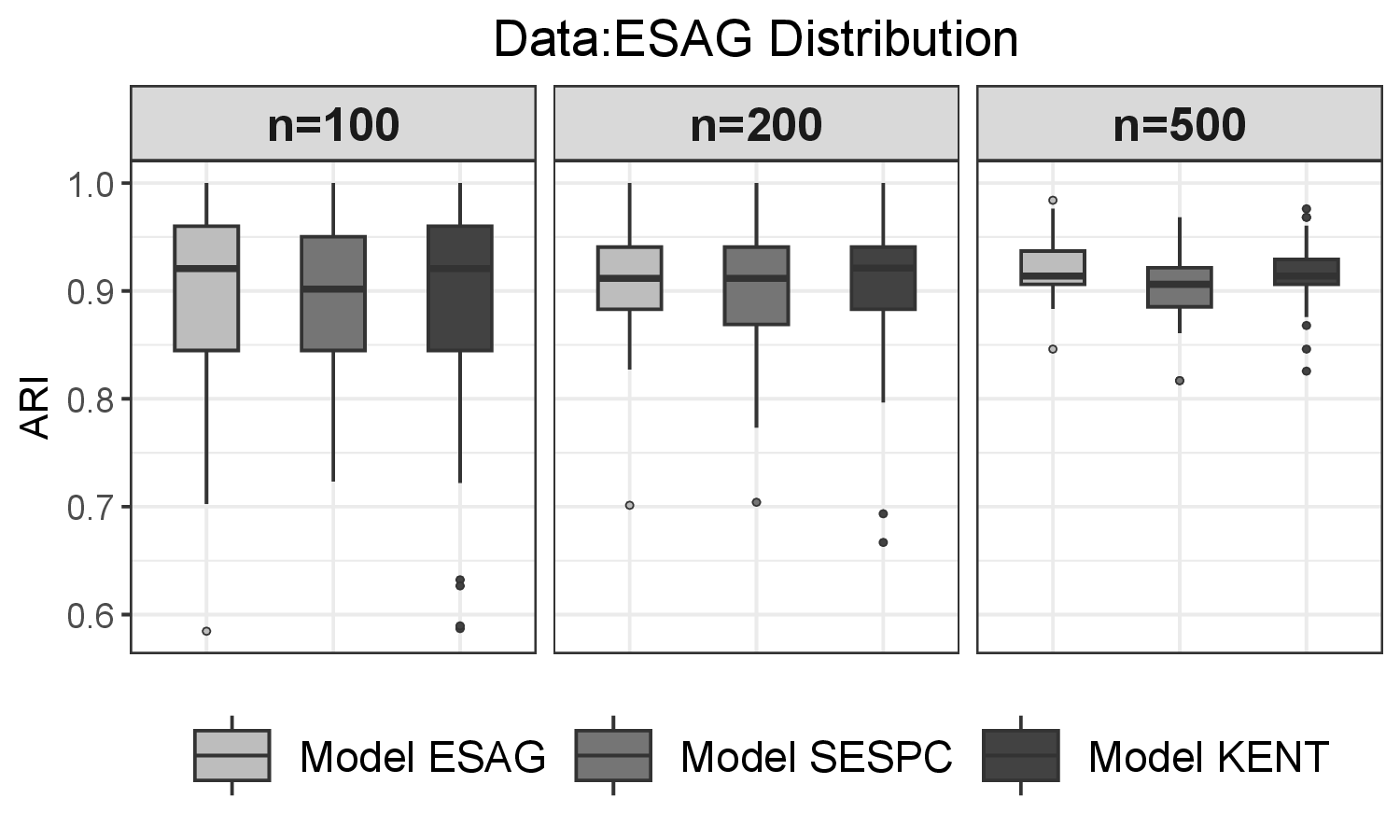} \\
\includegraphics[width=0.46\textwidth]{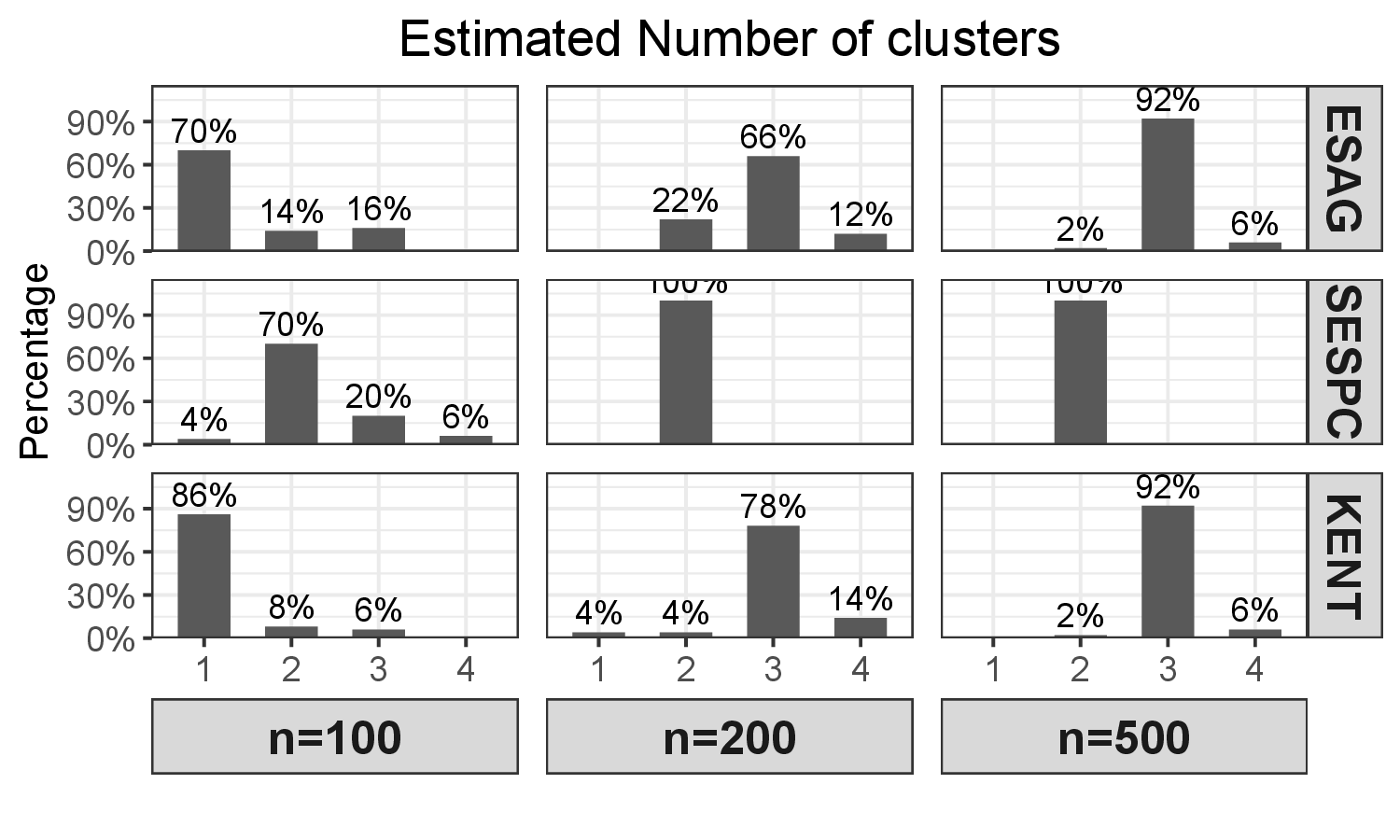}   \includegraphics[width=0.53\textwidth]{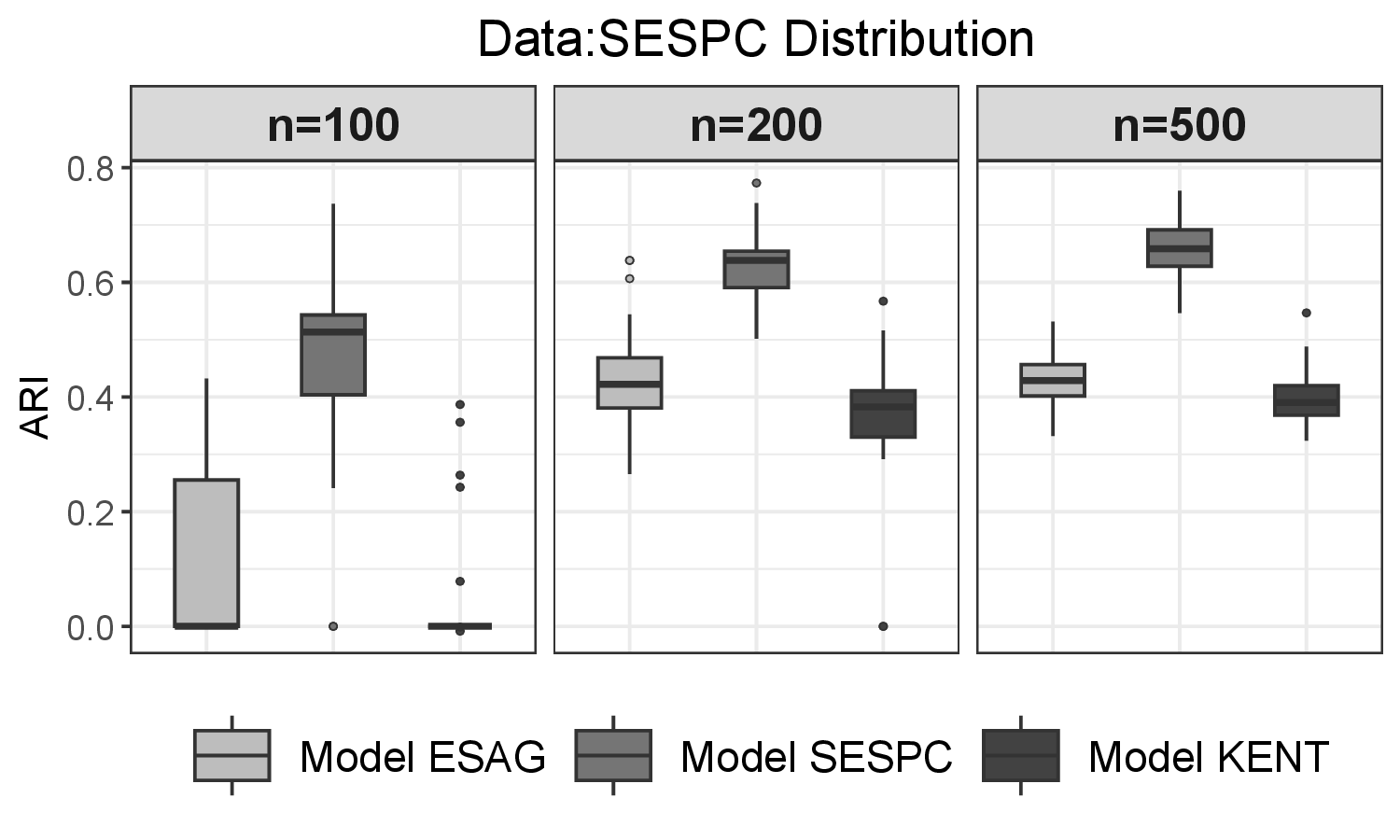}
\caption{ESAG and SESPC generated data with equal clusters: effect of $n$. $\bm{\gamma}=[(1,1)^\top,(1,1)^\top])$, $\omega=45^\circ$, $\bm{\mu_1'}=(-0.927,-0.282,0.243)^\top$, $\bm{\mu_2'}=(-0.886, -0.008,-0.464)^\top$ and various $\gamma$ combinations.}
\label{esag_effect_of_n}
\end{figure}

Figure \ref{esag_effect_of_n} illustrates the effect of sample size on the  performance of each model for $n \in \{100, 200, 500\}$, Without loss of generality we assume that, across all cases, the probabilities for an observation to belong at each cluster are the same. When data are generated from the ESAG distribution, the performance of both models remains relatively stable in terms of their ARI values  across all the investigated sample sizes, suggesting robustness with respect to $n$. In contrast, when the data arise from the SESPC distribution, the sample  size appears to have an effect on both the ability to recover the correct number of clusters and the models' performance as measured by the ARI. Specifically, an increase in sample size leads to higher with 
less variability ARI values, indicating improved clustering performance as $n$ grows.

\begin{figure}
\centering
\includegraphics[width=0.46\textwidth]{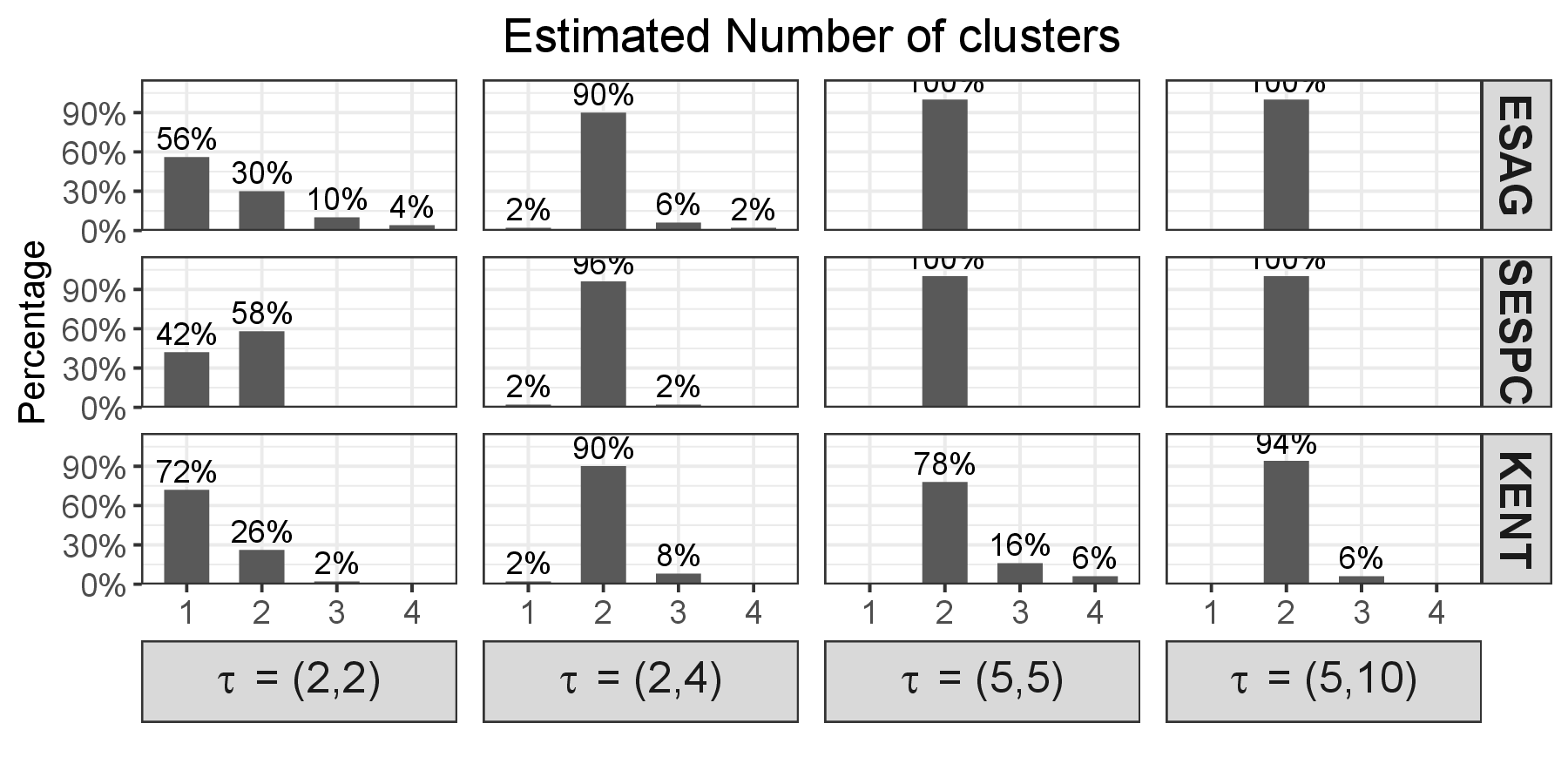}  
\includegraphics[width=0.53\textwidth]{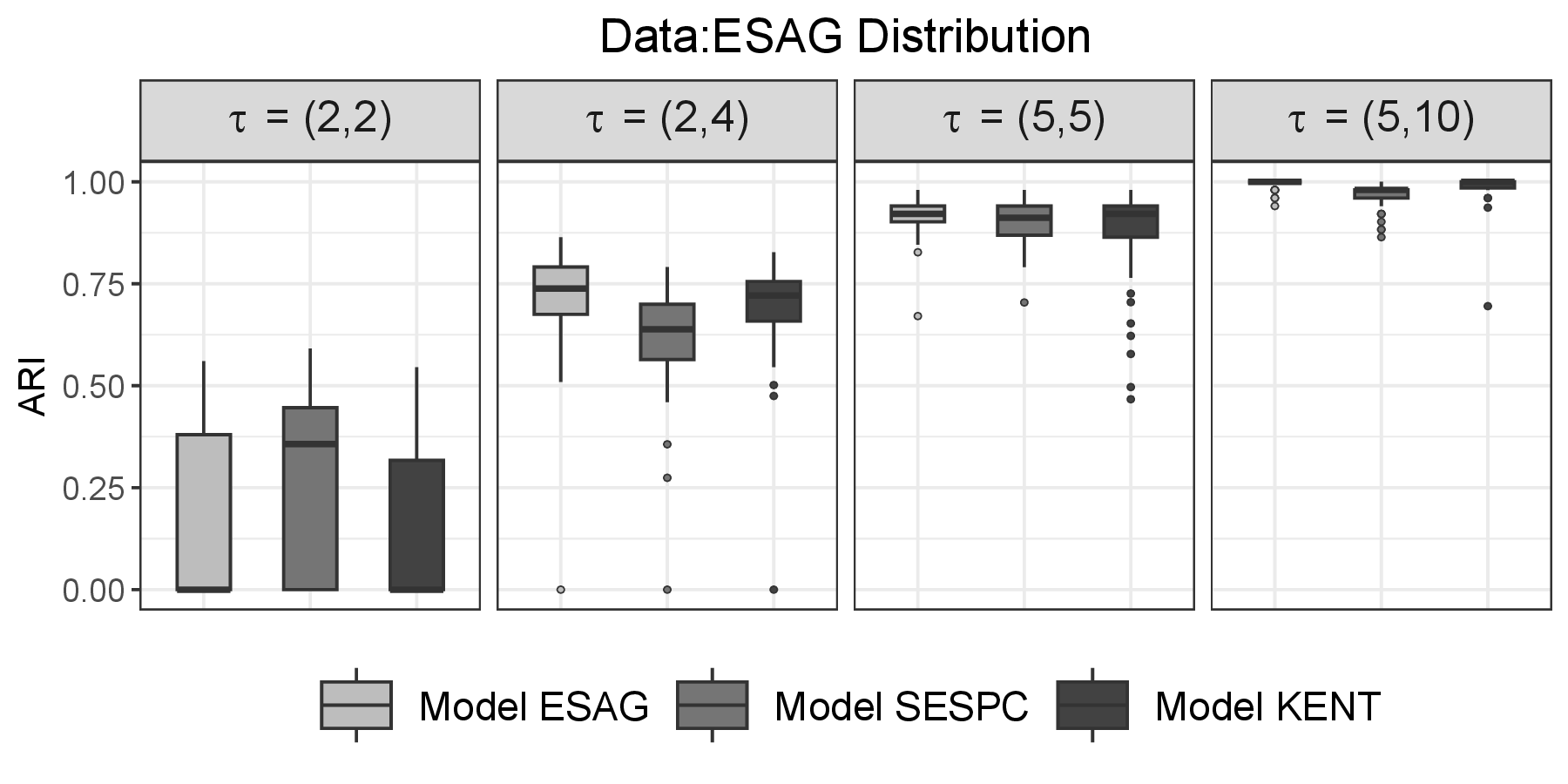} \\
\includegraphics[width=0.46\textwidth]{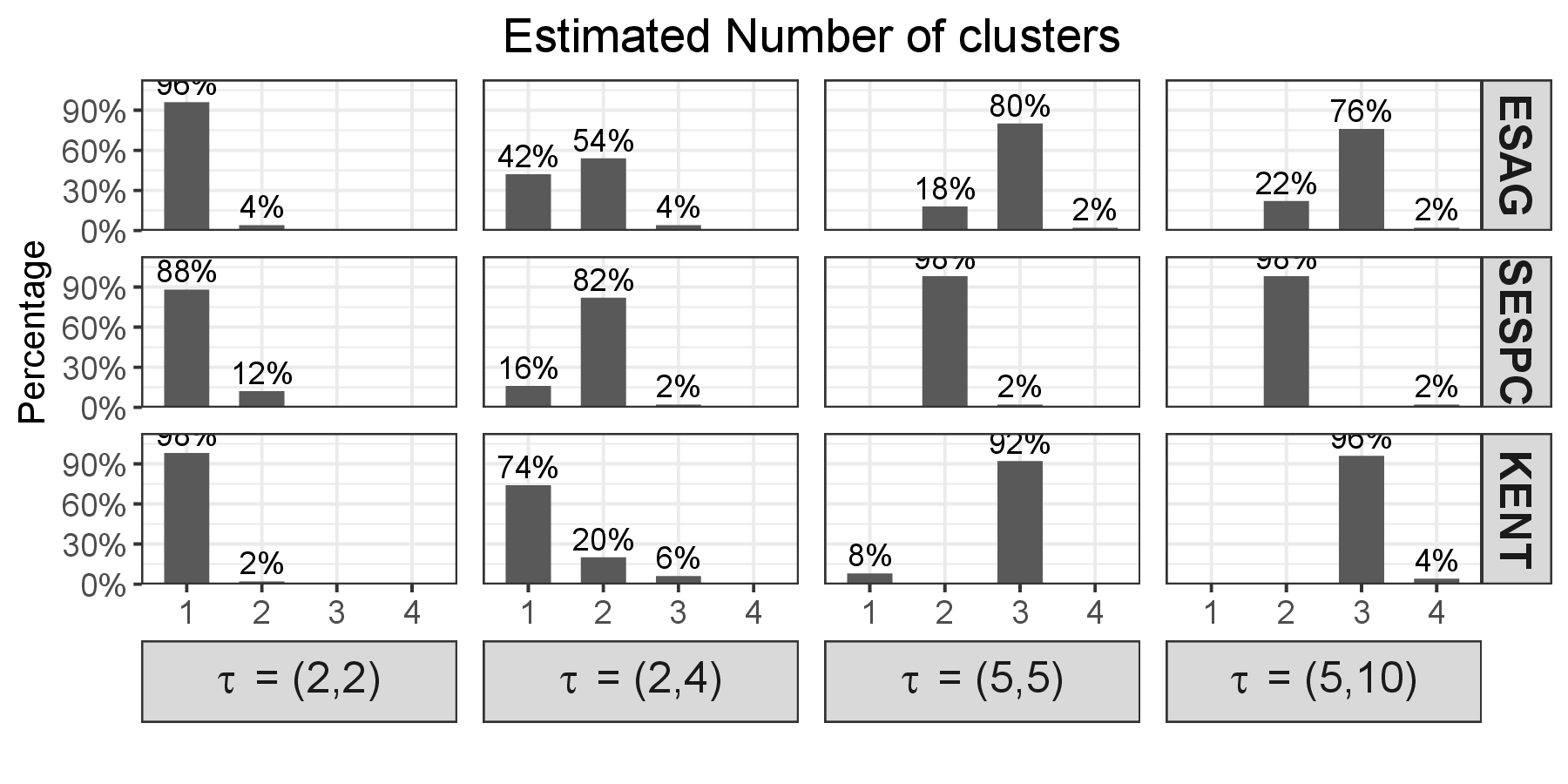}   
\includegraphics[width=0.53\textwidth]{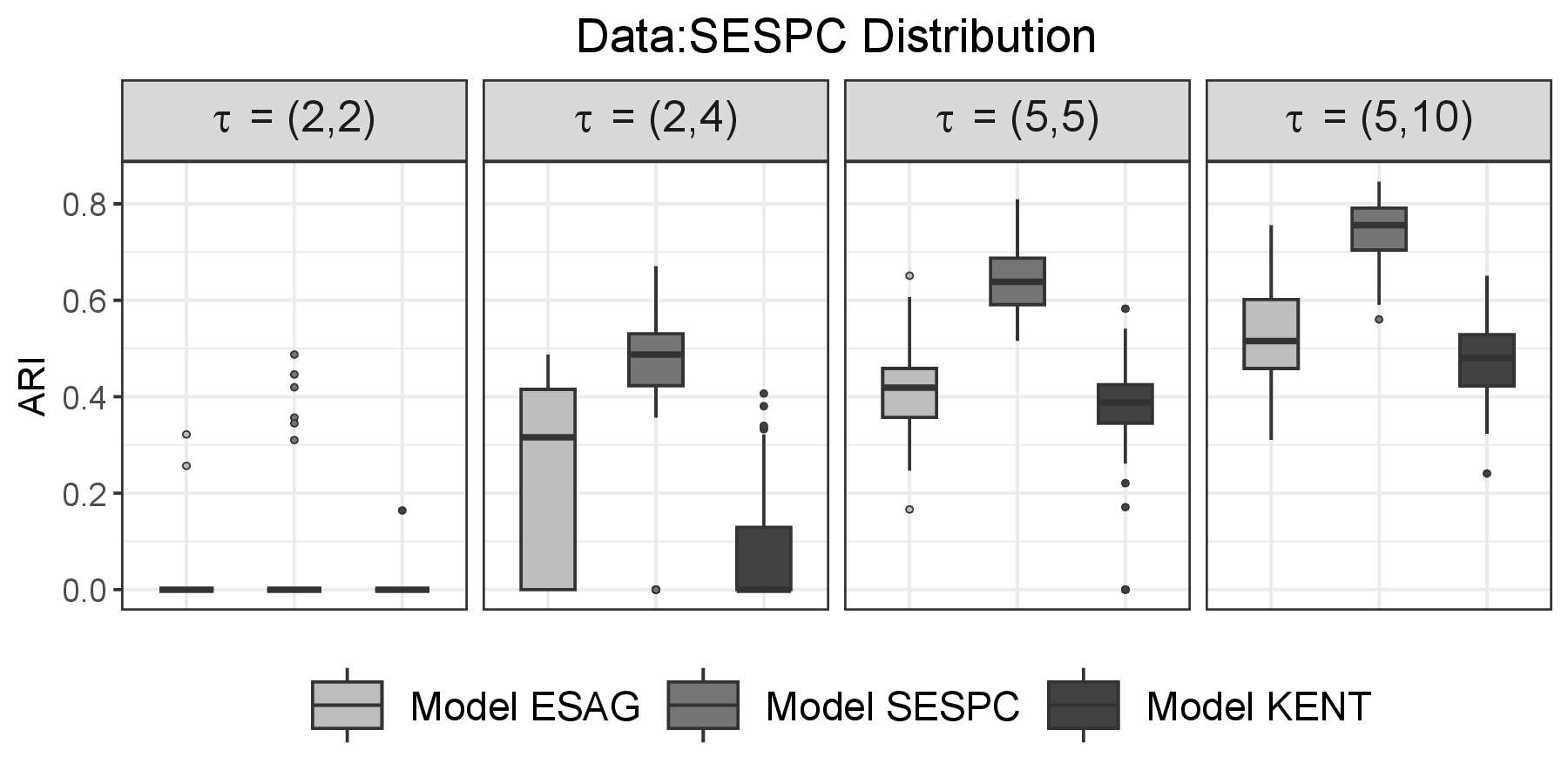} 
\caption{ESAG and SESPC generated data with equal clusters: effect of $\tau$. $n=200$, $\bm{\gamma}=[(1,1)^\top,(1,1)^\top]),\omega=45^\circ$, $\bm{\mu_1'}=(-0.927,-0.282,0.243)^\top$, $\bm{\mu_2'}=(-0.886, -0.008,-0.464)^\top$ and $(\tau_1,\tau_2)=[(2,2),(4,4),(5,5),(5,10)]$.}
\label{esag_effect_of_tau}
\end{figure}

Additional results are presented in Figure \ref{esag_effect_of_tau} regarding the effect of the concentration parameter $\bm{\tau}$. In general, small (large) values of $\tau_1$, $\tau_2$ result in small (large) spread in the corresponding clusters. This, as expected, affects the models' performance. In particular for small combination of $(\tau_1$, $\tau_2)=(2,2)$ both models do not perform well since are not able to detect the correct number of clusters in most of the cases. As the magnitude of the concentration increases the performance of each models is improved.

\subsection{Numerical analysis for hyper--spherical data}

\begin{figure}
\centering
\includegraphics[width=0.46\textwidth]{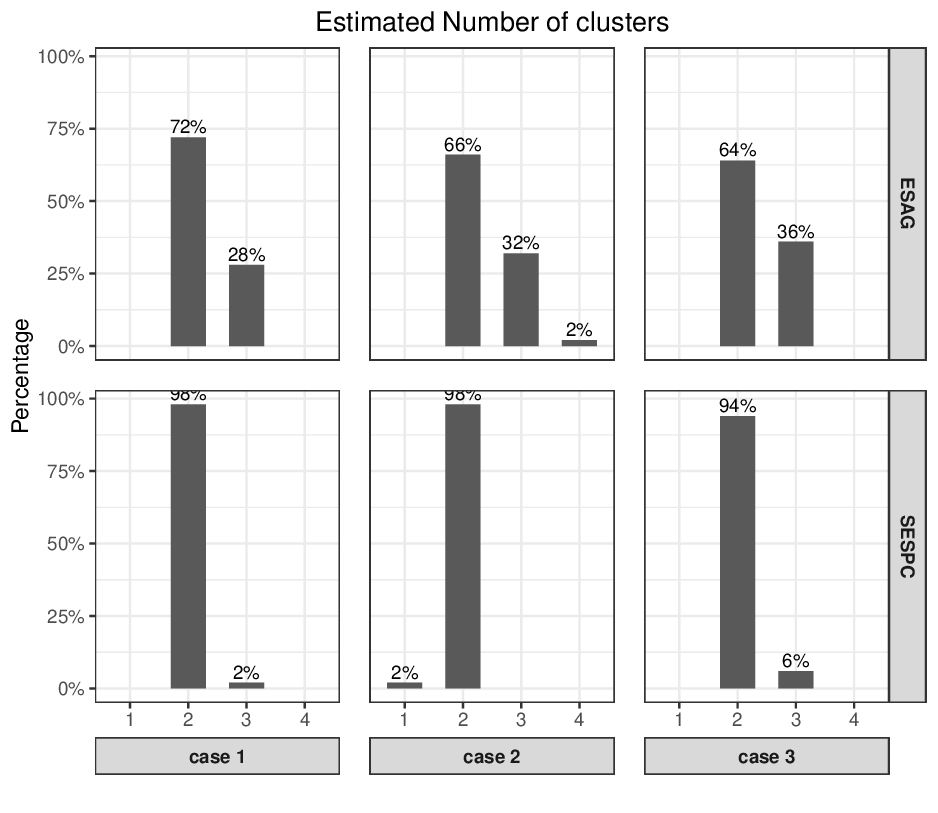} 
\includegraphics[width=0.53\textwidth]{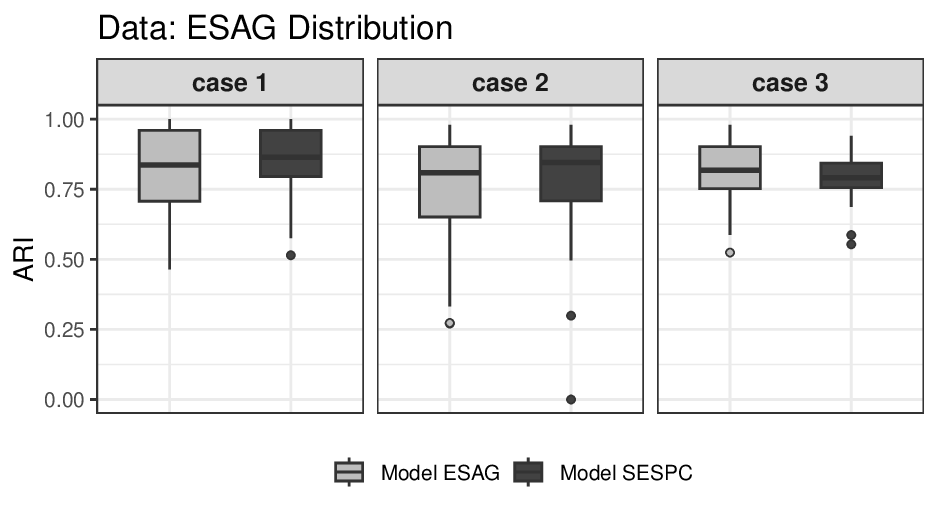} 
\caption{Hyper--spherical ESAG  generated data with equal clusters: $n=200$, $\omega=45^\circ$, The values of  $\bm{\gamma_1}, \bm{\gamma_2}$ and $\bm{\mu_1}, \bm{\mu_2}$ are presented in the Appendix.}
\label{hyper_n200_equal_prob_D_ESPC}
\end{figure}

\begin{figure}
\centering
\includegraphics[width=0.46\textwidth]{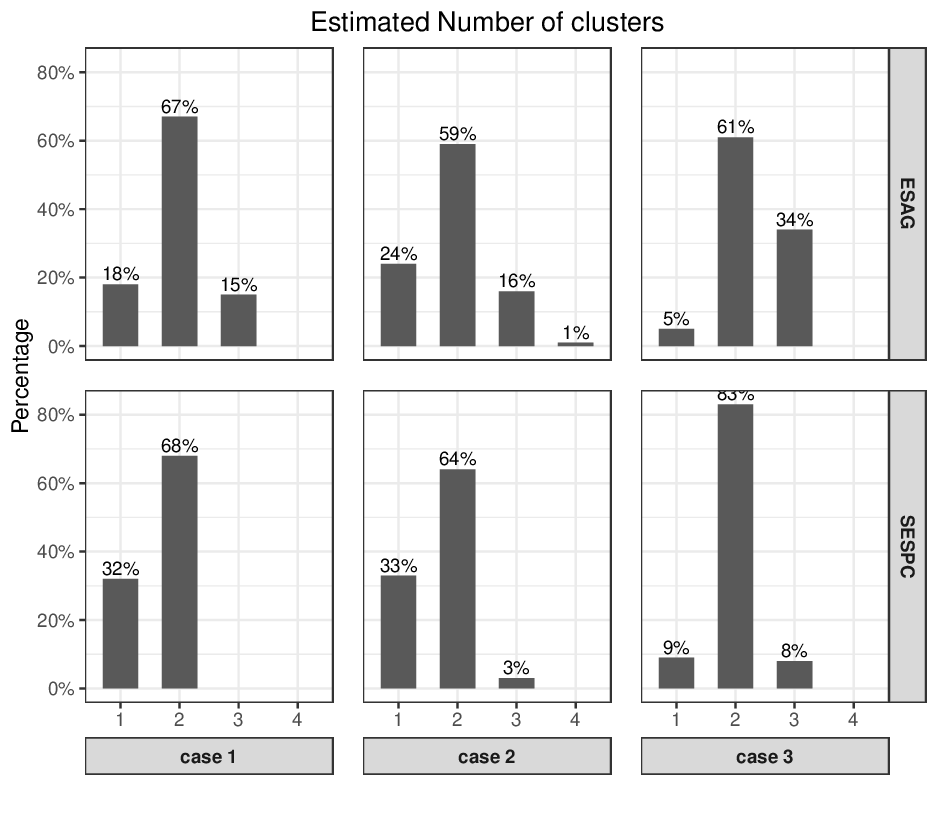} 
\includegraphics[width=0.53\textwidth]{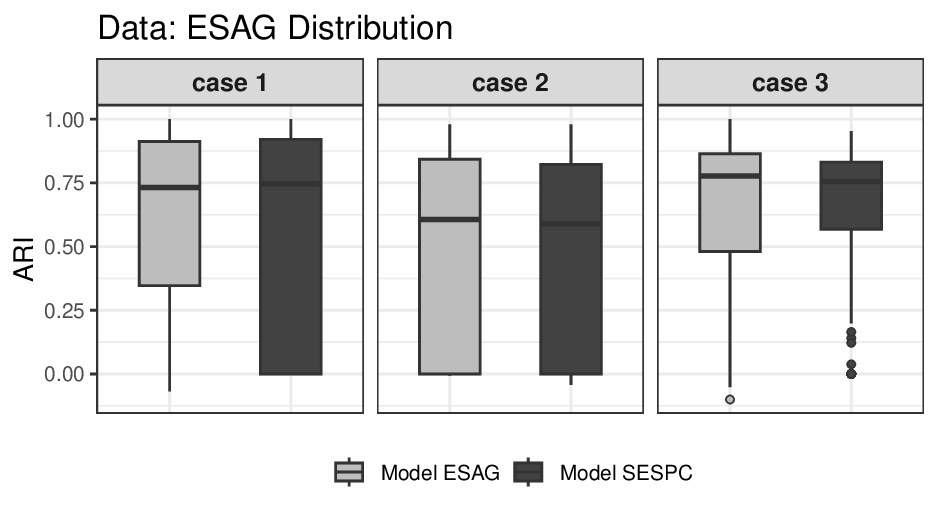} 
\caption{Hyper--spherical ESAG generated data with unequal clusters: $n=200$,  $\omega=45^\circ$, The values of  $\bm{\gamma_1}, \bm{\gamma_2}$ and $\bm{\mu_1}, \bm{\mu_2}$ are presented in the Appendix.}
\label{hyper_n200_unequal_prob_D_ESAG}
\end{figure}

\begin{figure}
\centering
\includegraphics[width=0.45\textwidth]{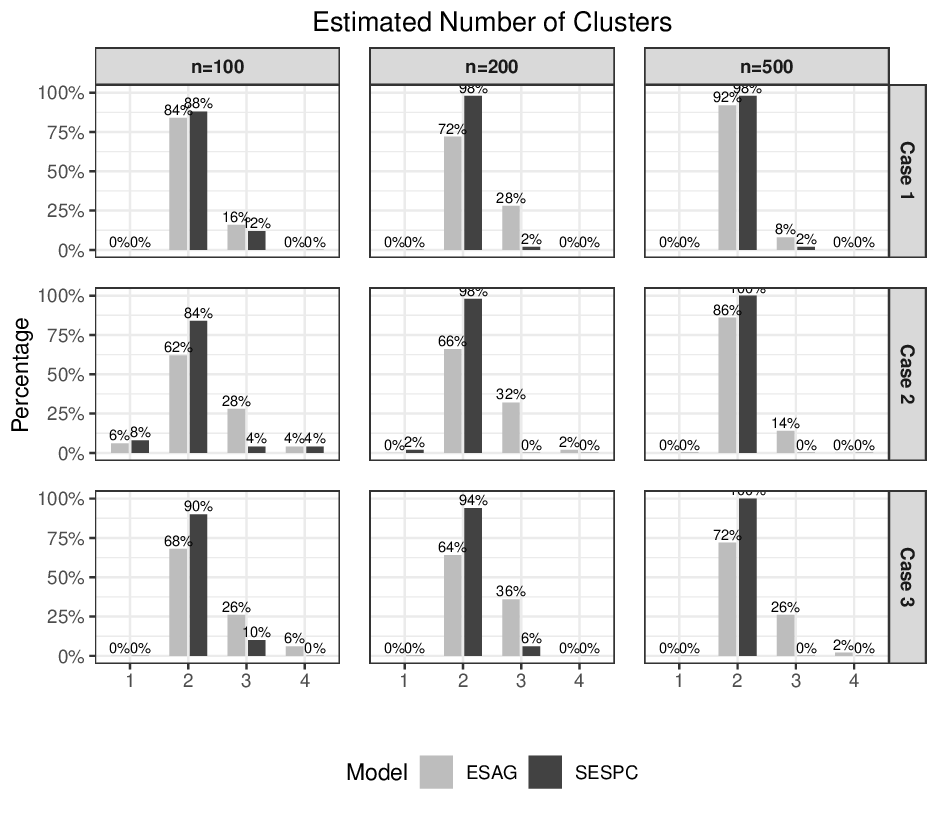} \includegraphics[width=0.45\textwidth]{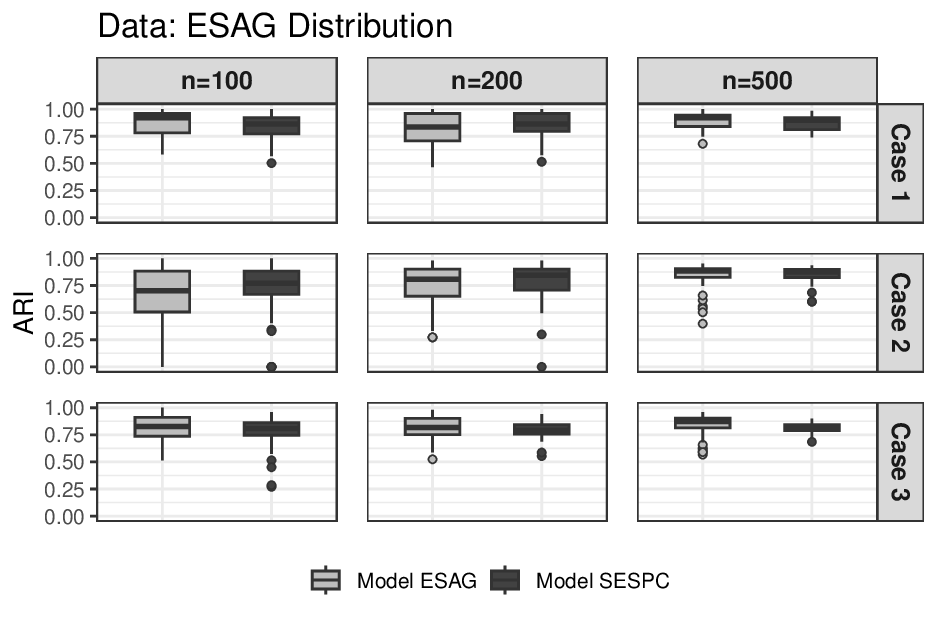}  \\
\caption{Hyper--spherical ESAG generated data with equal clusters: effect of $n$. $\omega=45^\circ$, The values of  $\bm{\gamma_1}, \bm{\gamma_2}$ and $\bm{\mu_1}, \bm{\mu_2}$ are presented in the Appendix.}
\label{hyper_esag_effect_of_n}
\end{figure}

The remainder of this section investigates the performance of the ESAG and SESPC models on hyper--spherical data of dimension $d=10$, considering the scenarios presented in Figure \ref{hyper_effect_of_gamma}. Specifically, we examine the effect of sample size, cluster size, and cluster configuration on the performance of each model. In Figure~\ref{hyper_effect_of_gamma}, 
analogously to the spherical case ($d=3$) examined in the previous section, equal cluster probabilities are assumed. From Figure \ref{hyper_esag_effect_of_n} it can be seen that, for $n=500$, both models have similar performance. Additionally, for $n=\{100,200\}$ the boxplots of the ARI values corresponding to the SESPC model have smaller variability and lie inside the boxplots which corresponds to the ESAG model. Lastly, the boxplots of ARIs regarding the performance of each model is presented under the assumption of having equal (Figure \ref{hyper_n200_equal_prob_D_ESPC}) and unequal (Figure \ref{hyper_n200_unequal_prob_D_ESAG}) cluster sizes. In both figures, we may observe that the relation between the performance of each model is roughly the same, with the difference that when having unequal clusters, as expected, the width of the boxplots is  larger.

\subsection{Computational cost} \label{sec:copcost}
\begin{figure}
\centering
\includegraphics[width=0.6\textwidth]{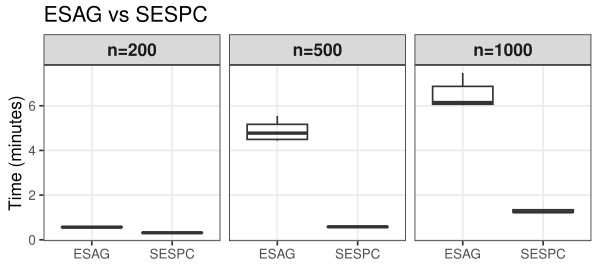}   
\caption{Box--plots of ESAG vs SESPC computational cost of searching for the optimal number of clusters.}
\label{fig:times}
\end{figure}

Figure \ref{fig:times} presents the computational cost for each model as a function of the sample size $n={200,500,1000}$. In particular, for each model, 10 iterations were considered and the required time (in minutes) to obtain the results was stored, by using a computer with Intel(R) Core(TM) i7--8550U CPU with 16 GB RAM without performing parallel computing(an option which is available for these computations). For each iteration, each algorithm searches for $K=1,\ldots,4$ clusters. The box--plots of  the needed time for each model for $n={200,500,1000}$  are presented in Figure \ref{fig:times}. It can be seen that, as expected, as the sample size increases the computational time also increases in both cases. Comparing the two models, it is evident that, regardless of the sample size, the SESPC model is faster. 

\section{Real data analysis} \label{sec:real}
Based on the analysis presented in the previous sections, the performance of the two models was evaluated across a range of scenarios involving simulated spherical and hyper--spherical datasets. The findings consistently demonstrate that the SESPC model exhibits strong performance when the underlying data arise from either the ESAG or the SESPC distribution, highlighting its adaptability across different frameworks. Having established the models' behaviour under controlled simulation settings, it would be interesting to assess the performance of both models on real data arising from various directional statistics application domains, including seismological recordings, as well as to explore their potential in clustering settings involving multivariate non--directional data. The first two datasets are located on the sphere, whereas the other two are Euclidean data that were normalized to become unit--vectors and reside on the hyper--sphere. The rest of this Section is outlined as follows: For each example, the true labelled data are visualized on the sphere, providing a reference for the underlying grouping structure. The clustering solutions produced by each model are then examined and compared, allowing us to evaluate  which model successfully recovers the true cluster configuration.

\subsection{Seismological data}
\begin{figure}[h!]
\centering
\begin{tabular}{cc}
\includegraphics[scale = 0.2]{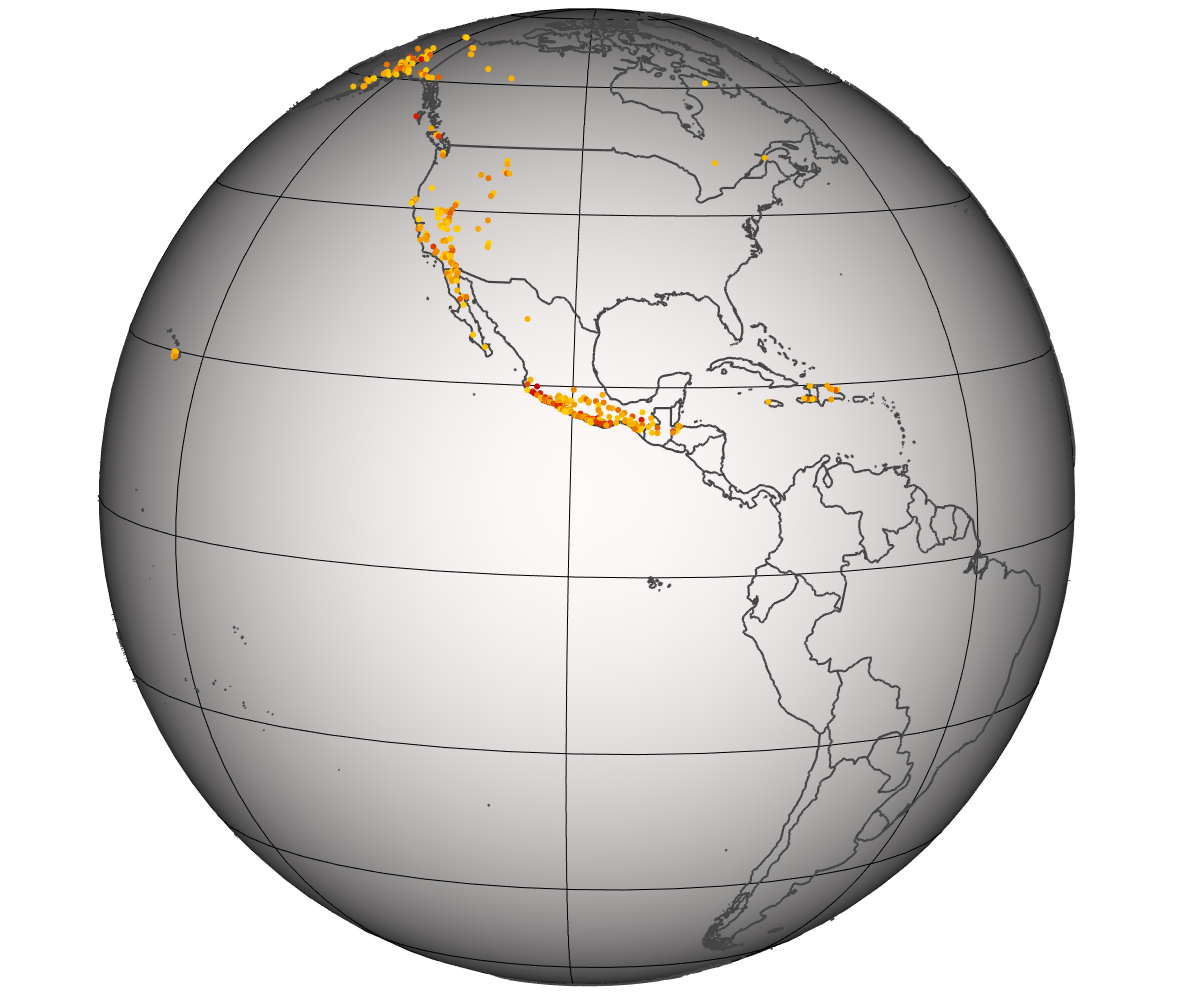}  &
\includegraphics[scale = 0.2]{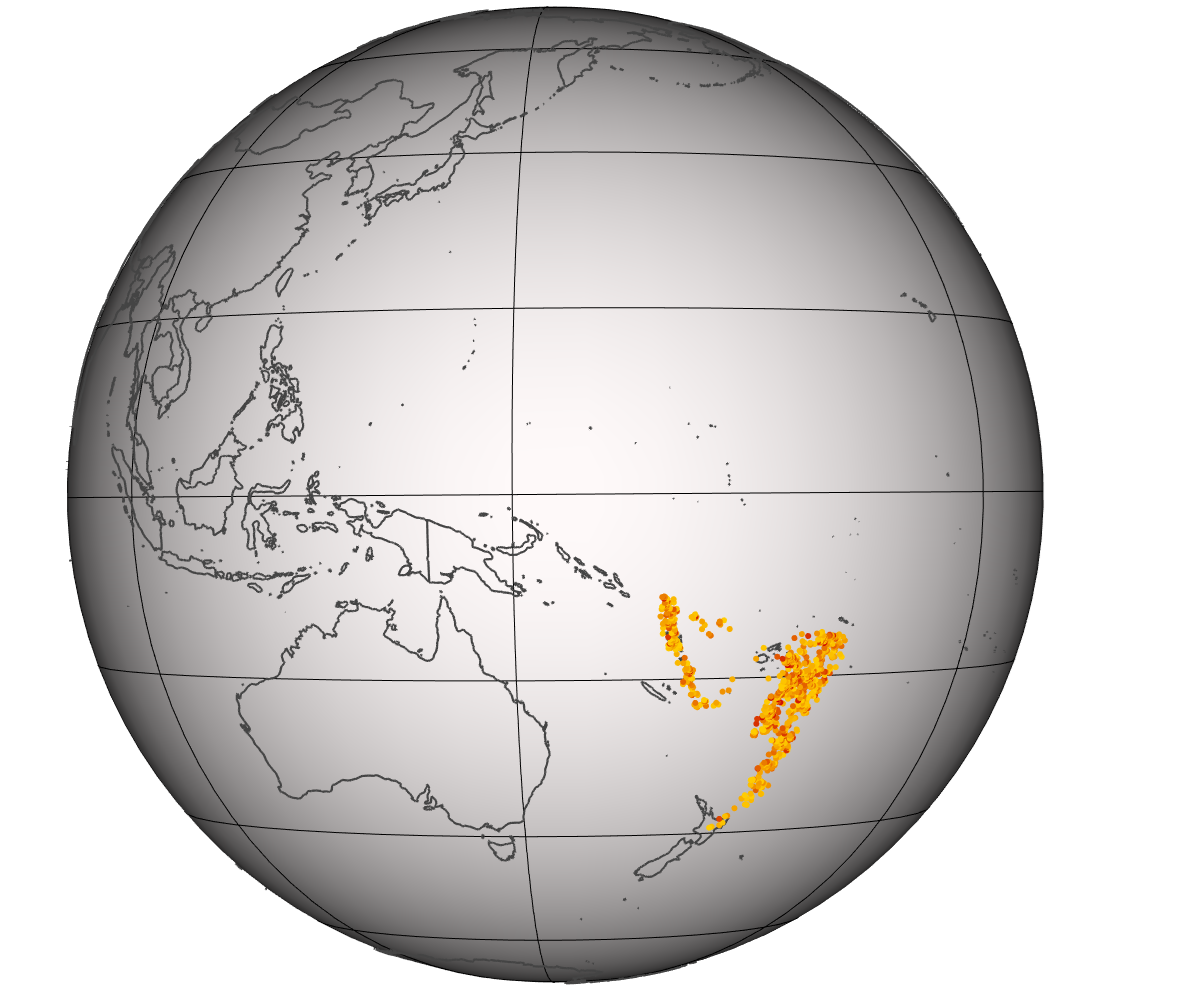} \\
(a) North America locations  &  (b) Fiji region locations
\end{tabular}
\caption{The locations of the earthquakes in North America and Fiji region. Different colors correspond to different levels of magnitude in Richter scale. \label{real} }
\end{figure}

The first two datasets contain locations of the earthquakes that took place (a) in the North American continent during 1900 and 2025\footnote{This dataset was downloaded from \href{https://earthquake.usgs.gov/fdsnws/event/1/query}{https://earthquake.usgs.gov}.} and (b) in a cube near Fiji since 1964\footnote{This dataset constitutes a subsample from a larger dataset of containing 5,000 observations and is available by default in \textit{R} by the names \textit{quakes}.} For North America we collected the earthquakes whose magnitude was 5 or higher in the Richter scale, whereas for the Fiji region the magnitude of the earthquakes was of 4 Richter or higher. North America contributes with 366 locations from Canada, Dominican Republic, Guatemala, Haiti, Jamaica, Mexico, and USA, with the last two countries containing the 87.43\% of the observations. The Fiji region consists of 1,000 locations from New Zealand, Solomon Islands and Vanuatu, but the vast majority (98.3\%) refers to earthquakes that happened offshore. Figure \ref{real} shows the locations of the earthquakes.

\subsubsection{North America}
\begin{figure}[h!]
\centering
\begin{tabular}{cc}
\includegraphics[scale = 0.29]{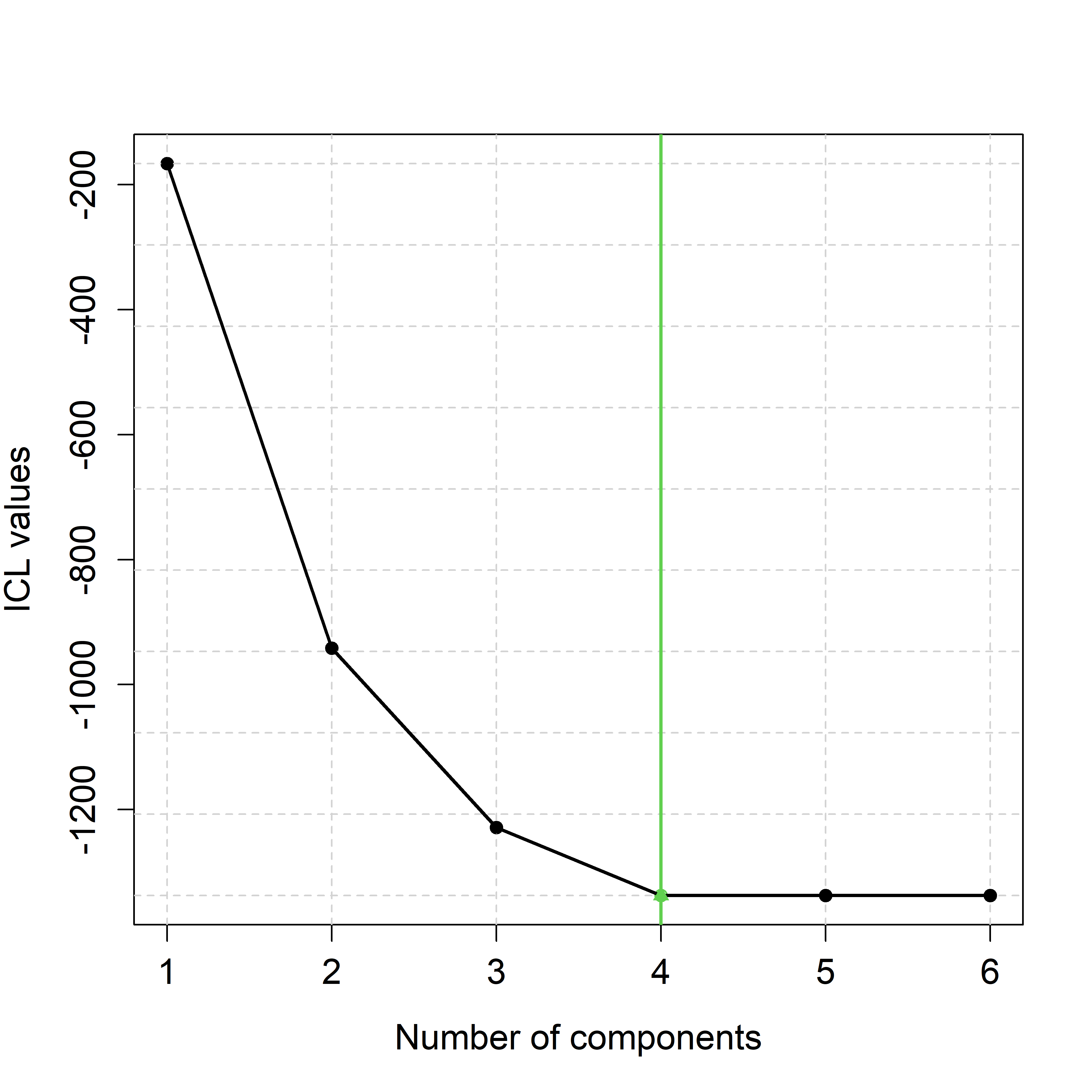}  &
\includegraphics[scale = 0.29]{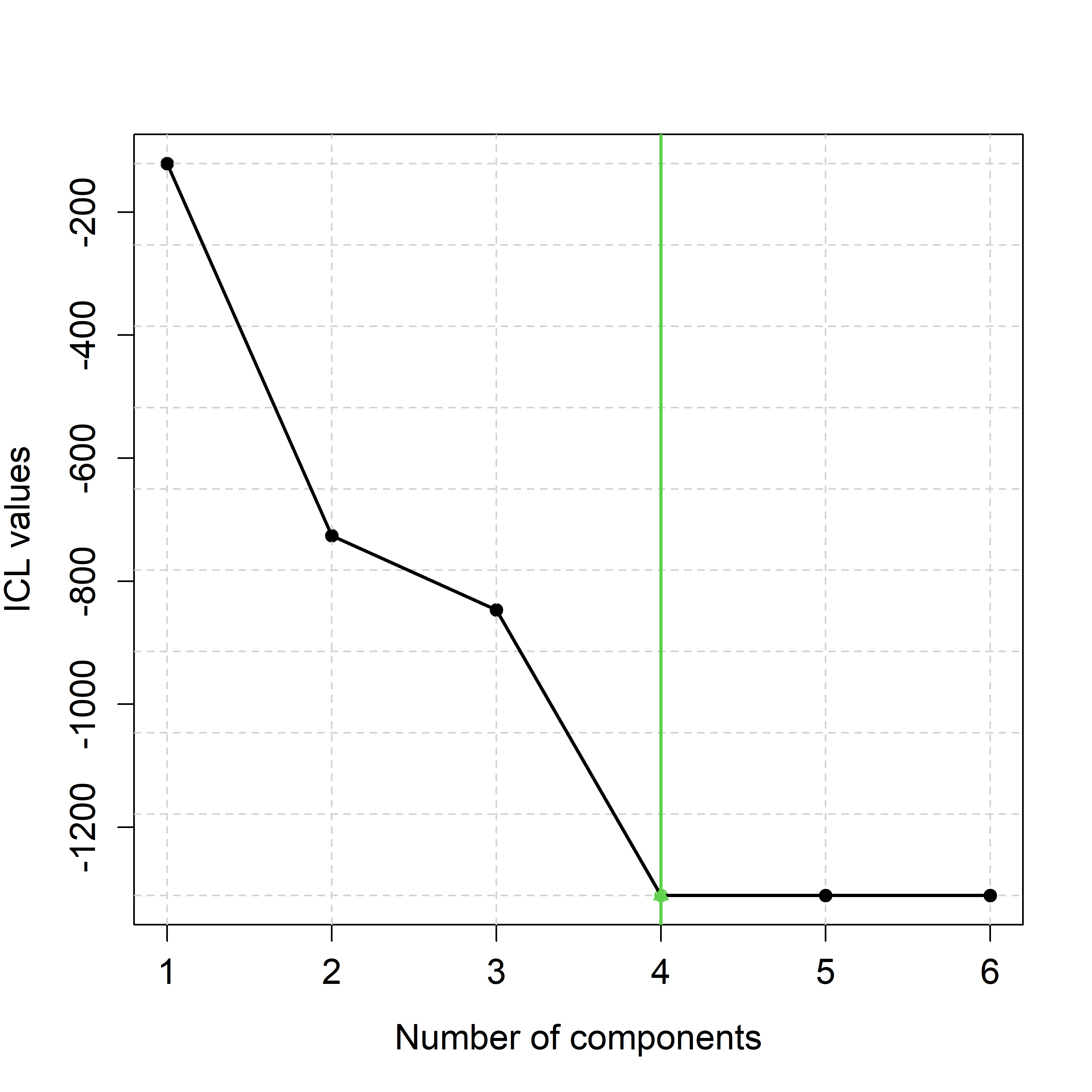} \\
(a) ICL for SESPC mixtures &  (b) ICL for ESAG mixtures \\
\includegraphics[scale = 0.25]{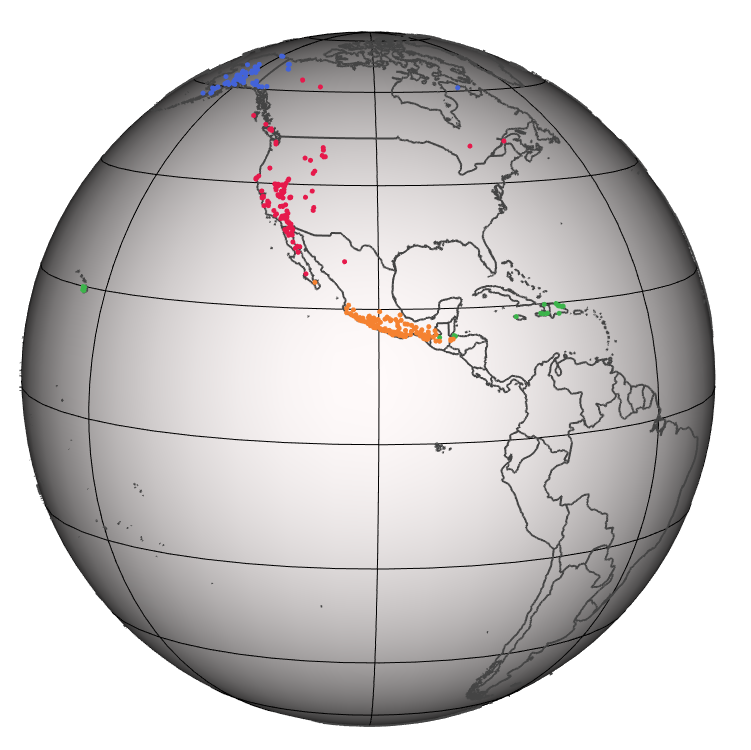}  &
\includegraphics[scale = 0.25]{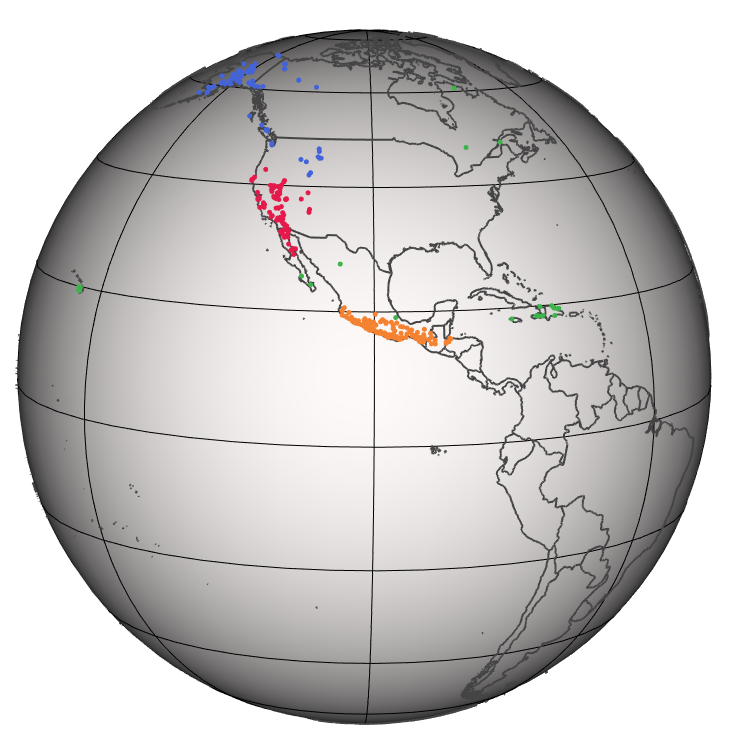} \\
(c) Spherical plot of SESPC mixtures  &  (d) Spherical plot of ESAG mixtures \\
\includegraphics[scale = 0.25]{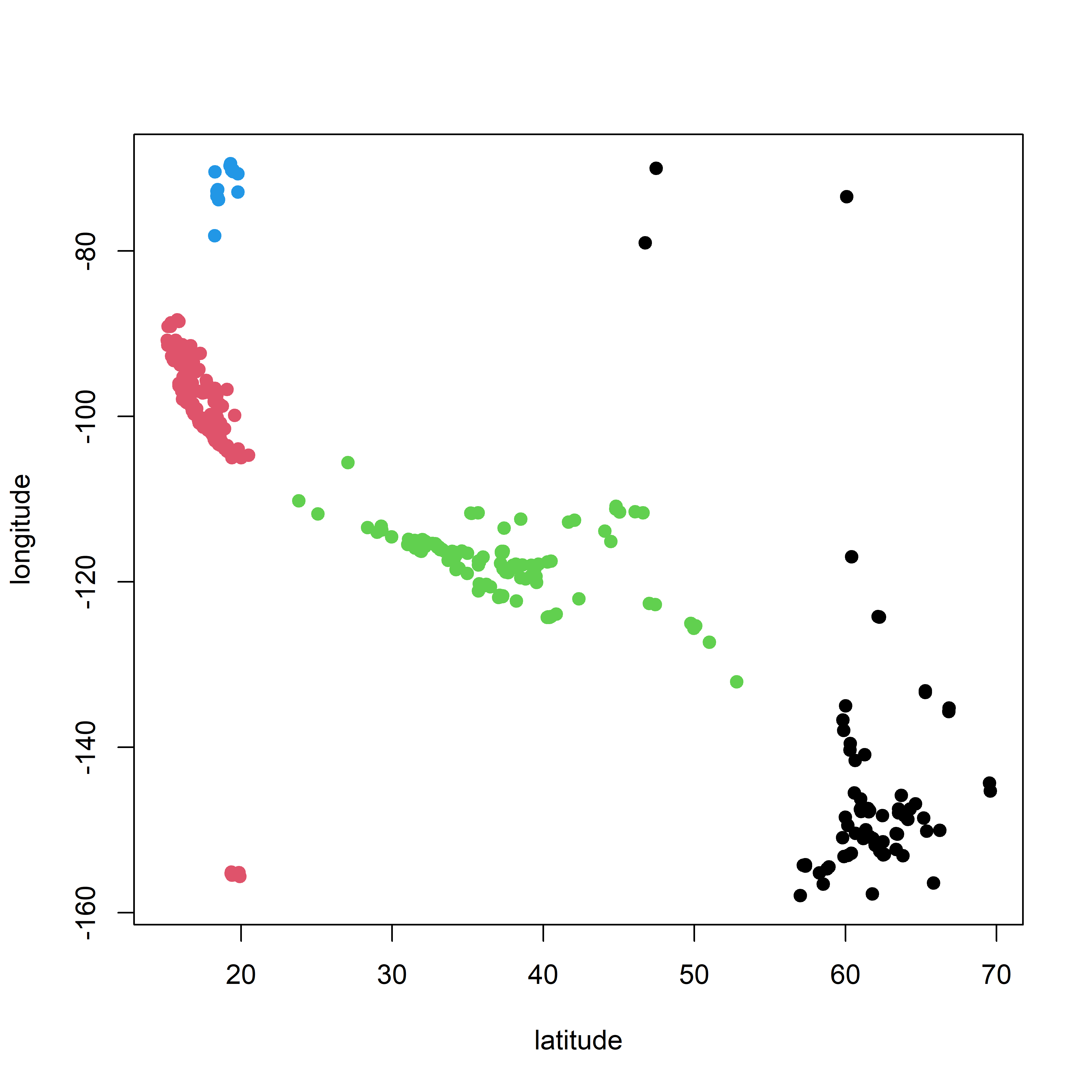}  &
\includegraphics[scale = 0.25]{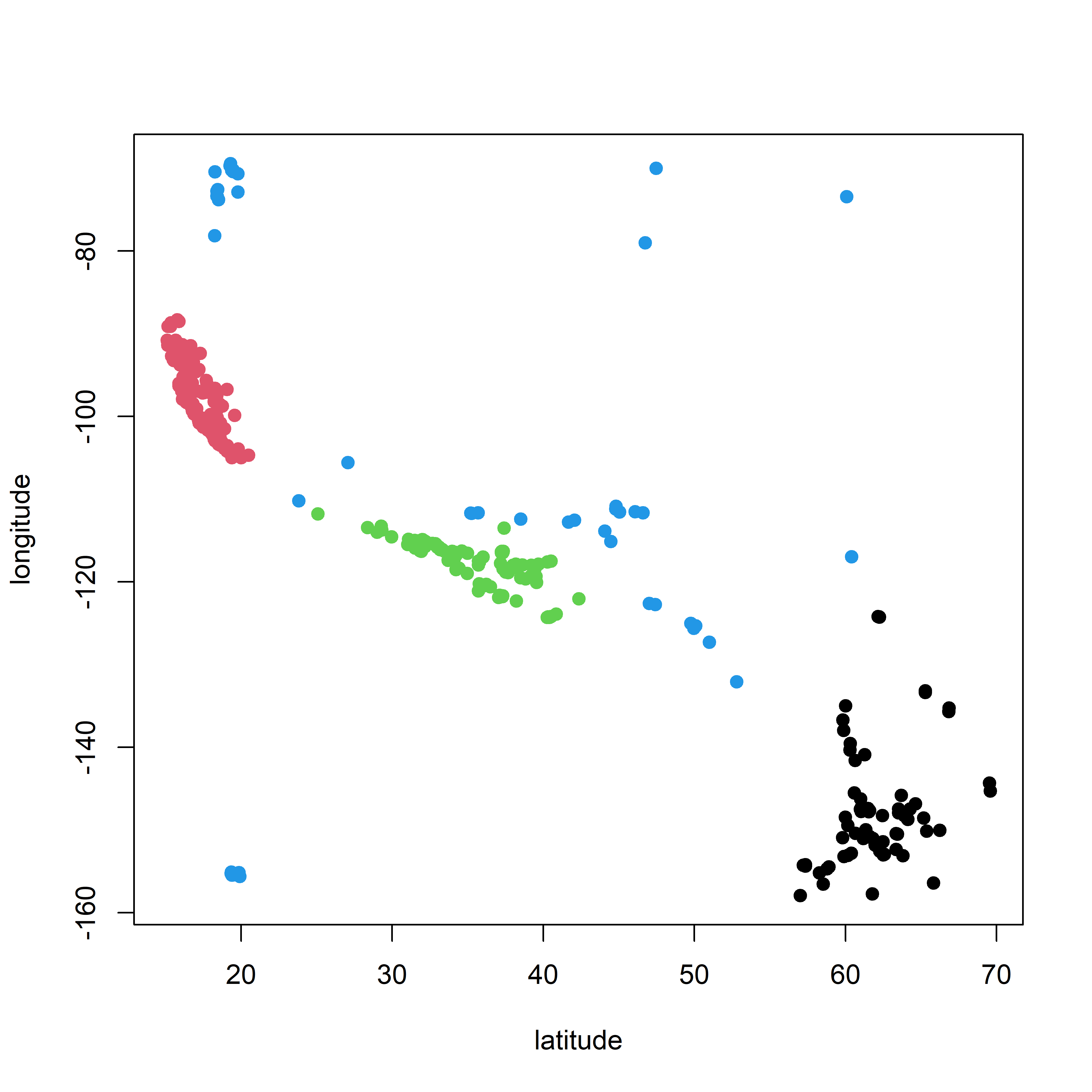} \\
(e) 2D plot of SESPC mixtures  &  (f) 2D plot of ESAG mixtures \\
\includegraphics[scale = 0.25]{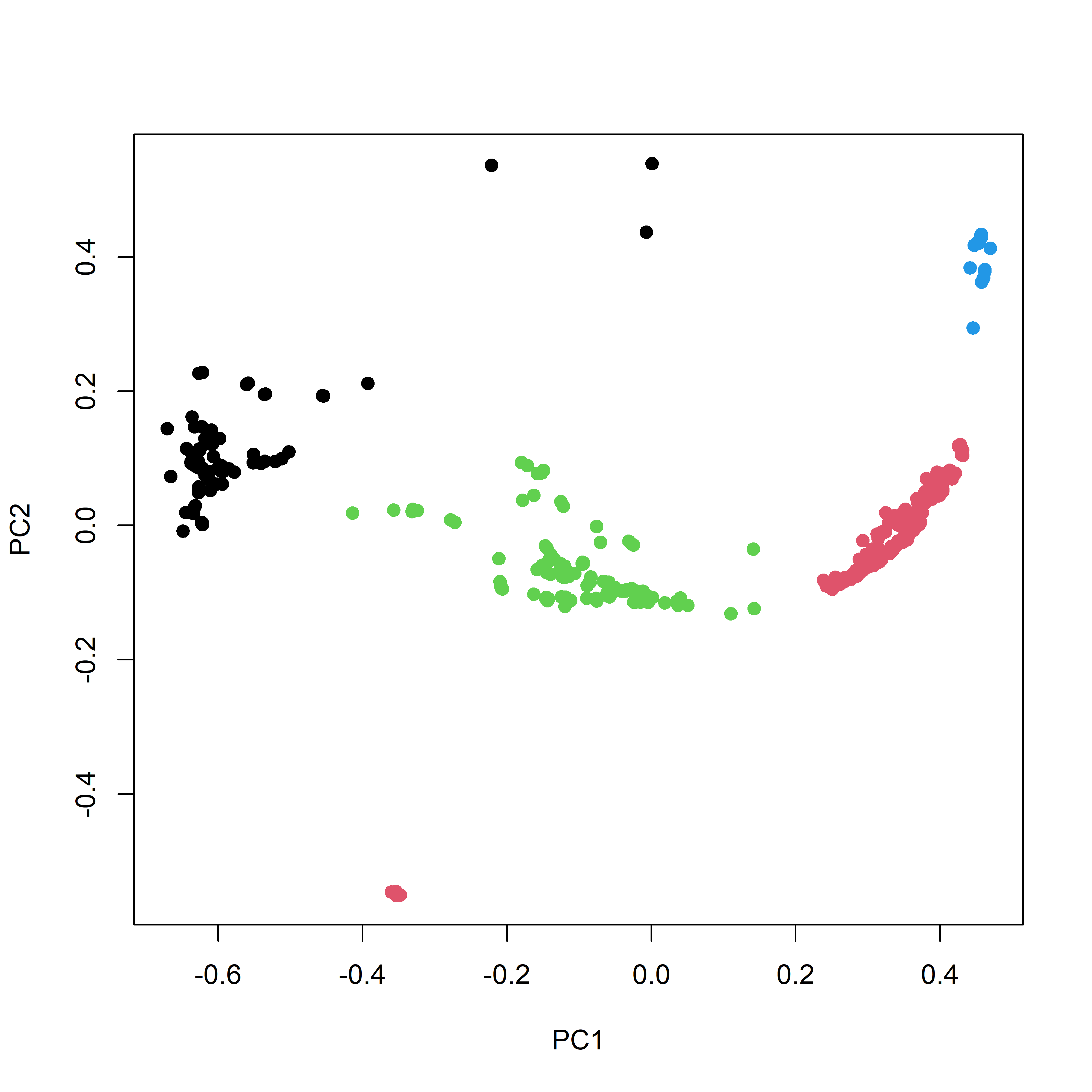}  &
\includegraphics[scale = 0.25]{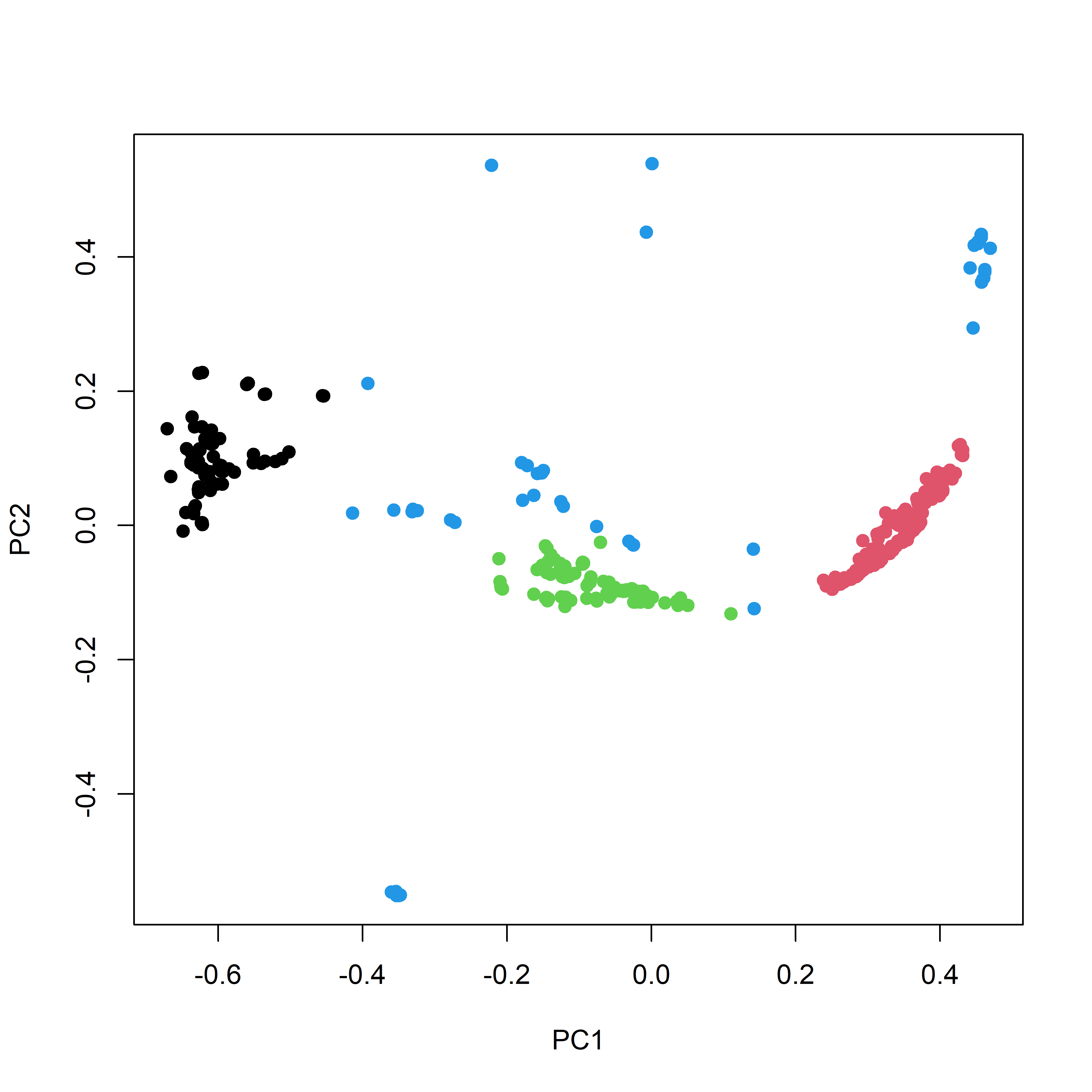} \\
(g) PCA plot of SESPC mixtures  &  (h) PCA plot of ESAG mixtures \\
\end{tabular}
\caption{North America: ICL plots for the two mixtures and locations of the earthquakes coloured according to their estimated cluster. \label{north} }
\end{figure}

As it was previously stated our goal is to investigate the performance of each model in real life dataset. Regarding the first dataset to be investigated, for both models, we searched for the optimal number of components $K$, using $K=1,\ldots,6$. Regarding the computational cost between the two approaches no significant difference occurred. In particular, fitting the SESPC, the search required 4.75 seconds, whereas the ESAG required 6.51 seconds. Figures \ref{north}(a)--\ref{north}(b) present the ICL values for the mixtures of both distributions using the selected range of $K$, indicating that both mixtures agree that 4 clusters are enough. Figures \ref{north}(c)--\ref{north}(d) visualize, on the sphere, the 4 clusters identified by each mixture of distributions. Evidently, the two distributions agree in the  blue and green clusters, but disagree in the red and orange clusters. Hawaii is clustered as green color in both models. Notable differences also occur with some western and eastern USA locations, as indicated by the different colours. It should be highlighted the fact that  that the SESPC mixtures have discriminated the locations almost perfectly, they are linearly separable, whereas the ESAG mixtures have created overlapping clusters that are not linearly separable. 

An alternative  visualization of the results is illustrated in Figures \ref{north}(e)--\ref{north}(f), where the locations(latitude and longitude) are placed in a 2D plot. Note that, distances in the 2D plot are not geodesic distances, two points that look far apart may be close on the sphere and vice versa. Since the clustering was done in 3D Cartesian space (or on the sphere directly), a visually comparison of the latitude/longitude coordinates can make a geometrically good clustering look bad and vice versa. For this reason we also projected the data onto the first two principal components of the Cartesian coordinates. Figure \ref{north}(g)--\ref{north}(h) shows the PCA projected data and the results are nearly the same, the 4 groups are nearly linearly separable using the SESPC mixtures.

We repeated the analysis creating a concomitant model, using the magnitude of the earthquakes as a covariate to link to the mixing probabilities. We did not attempt the mixture of regressions model simply because the magnitude is affected by the locations and not vice versa, but it seemed natural to affect the mixing probabilities. The results were nearly the same, the effect of the magnitude was shown to be rather weak. 

\subsubsection{Fiji region}
\begin{figure}[h!]
\centering
\begin{tabular}{cc}
\includegraphics[scale = 0.3]{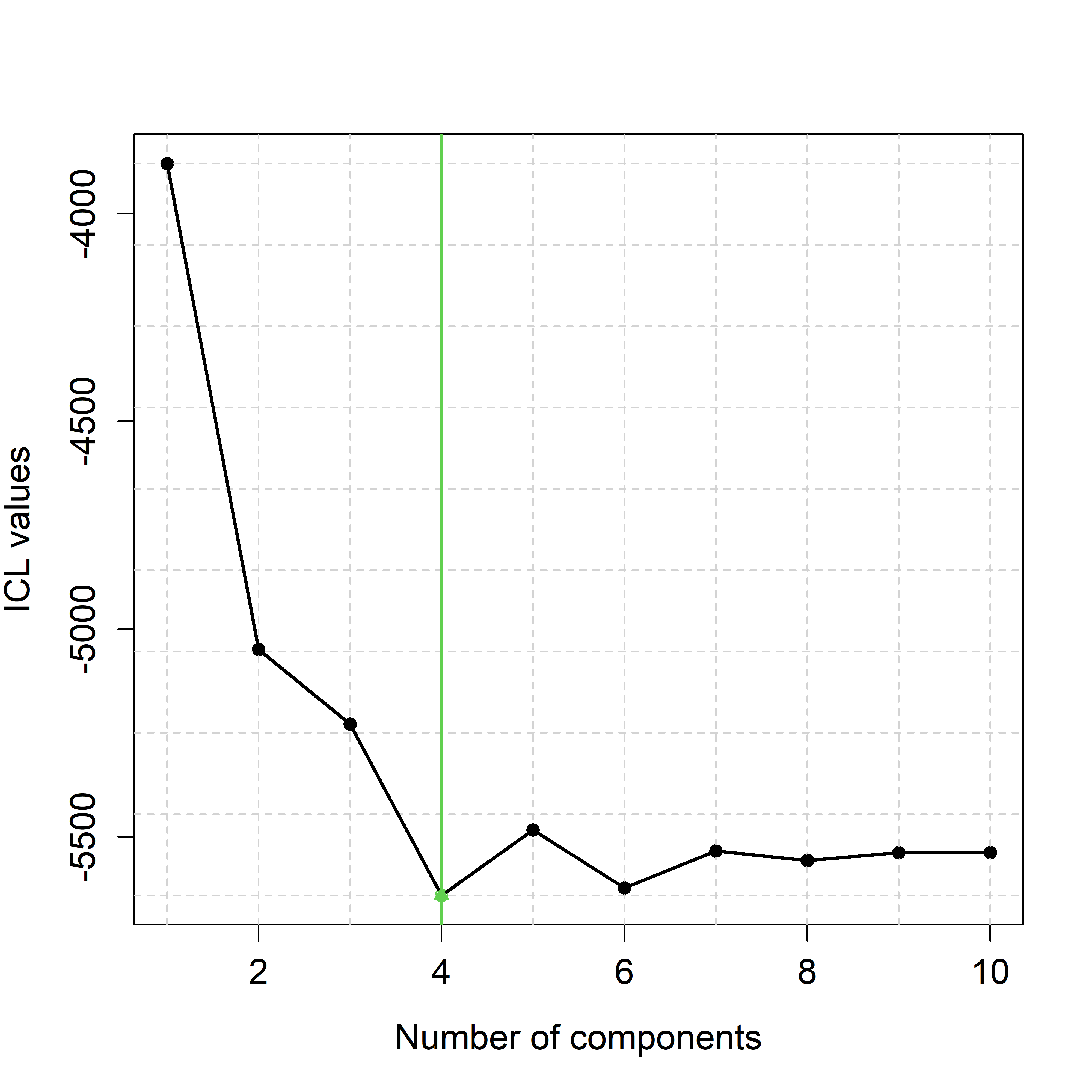}  &
\includegraphics[scale = 0.3]{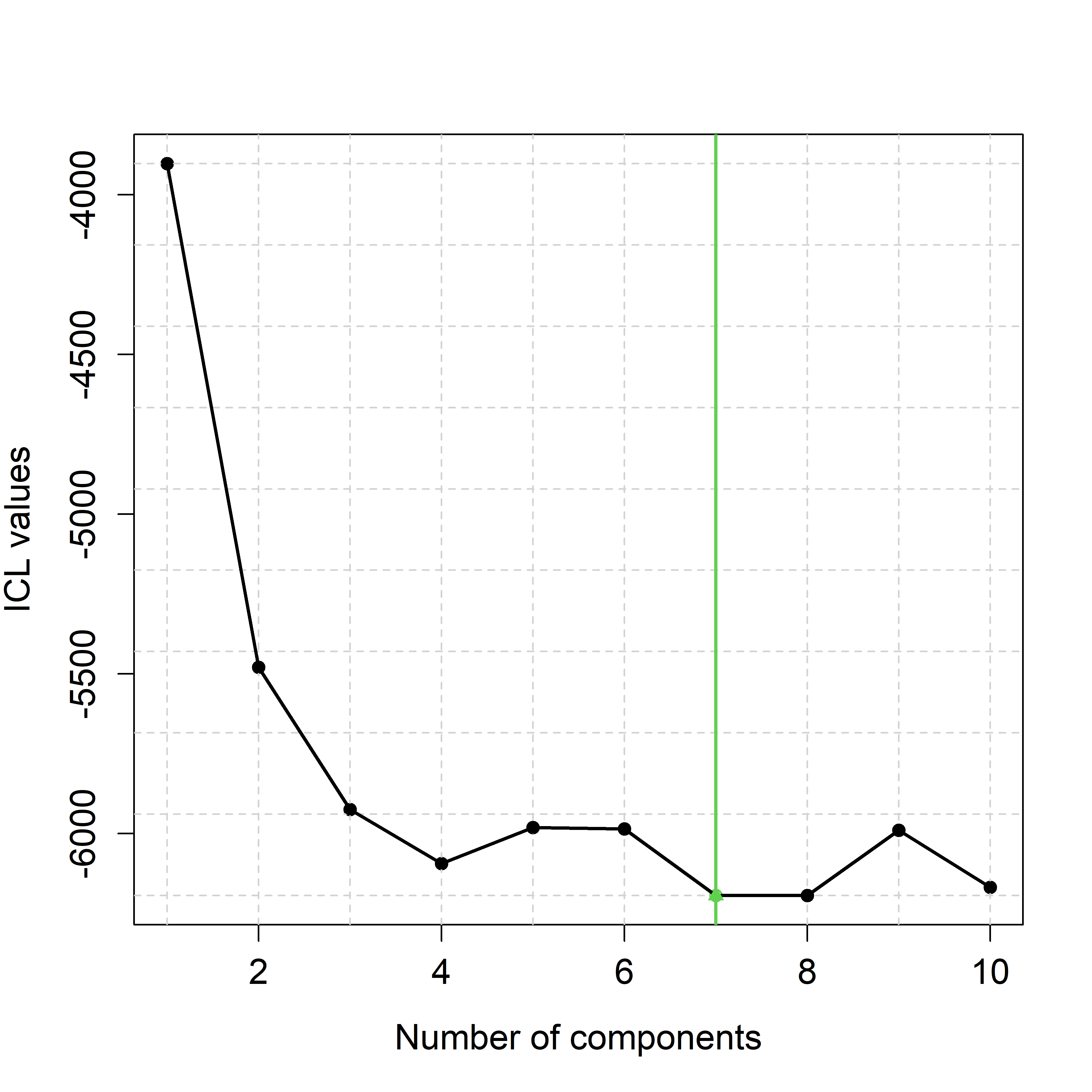} \\
(a) ICL for SESPC mixtures &  (b) ICL for ESAG mixtures \\
\includegraphics[scale = 0.35]{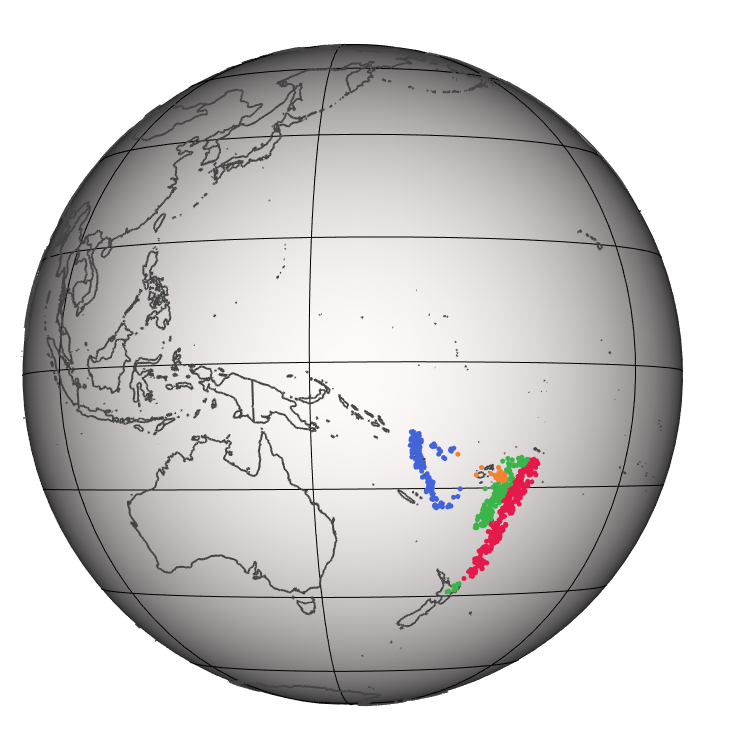}  &
\includegraphics[scale = 0.35]{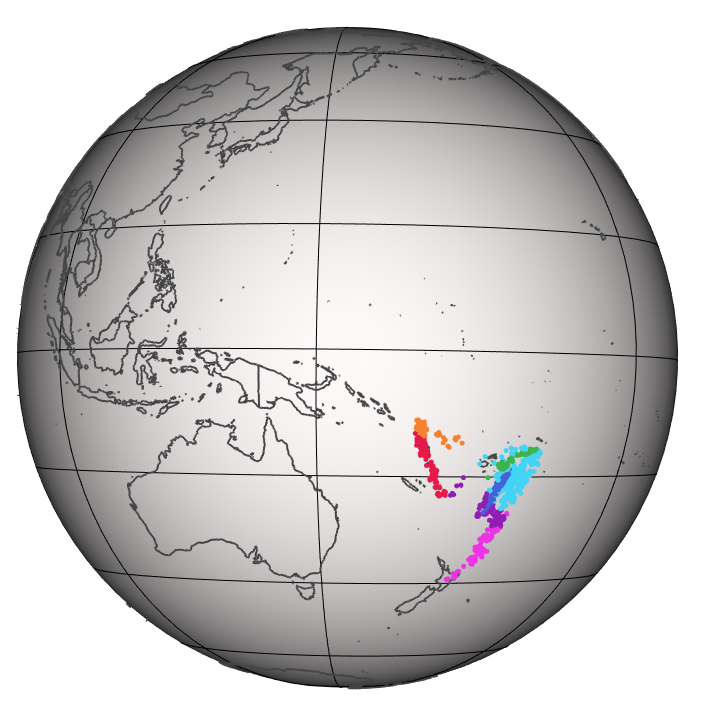} \\
(c) Spherical plot of SESPC mixtures  &  (d) Spherical plot of ESAG mixtures \\
\includegraphics[scale = 0.3]{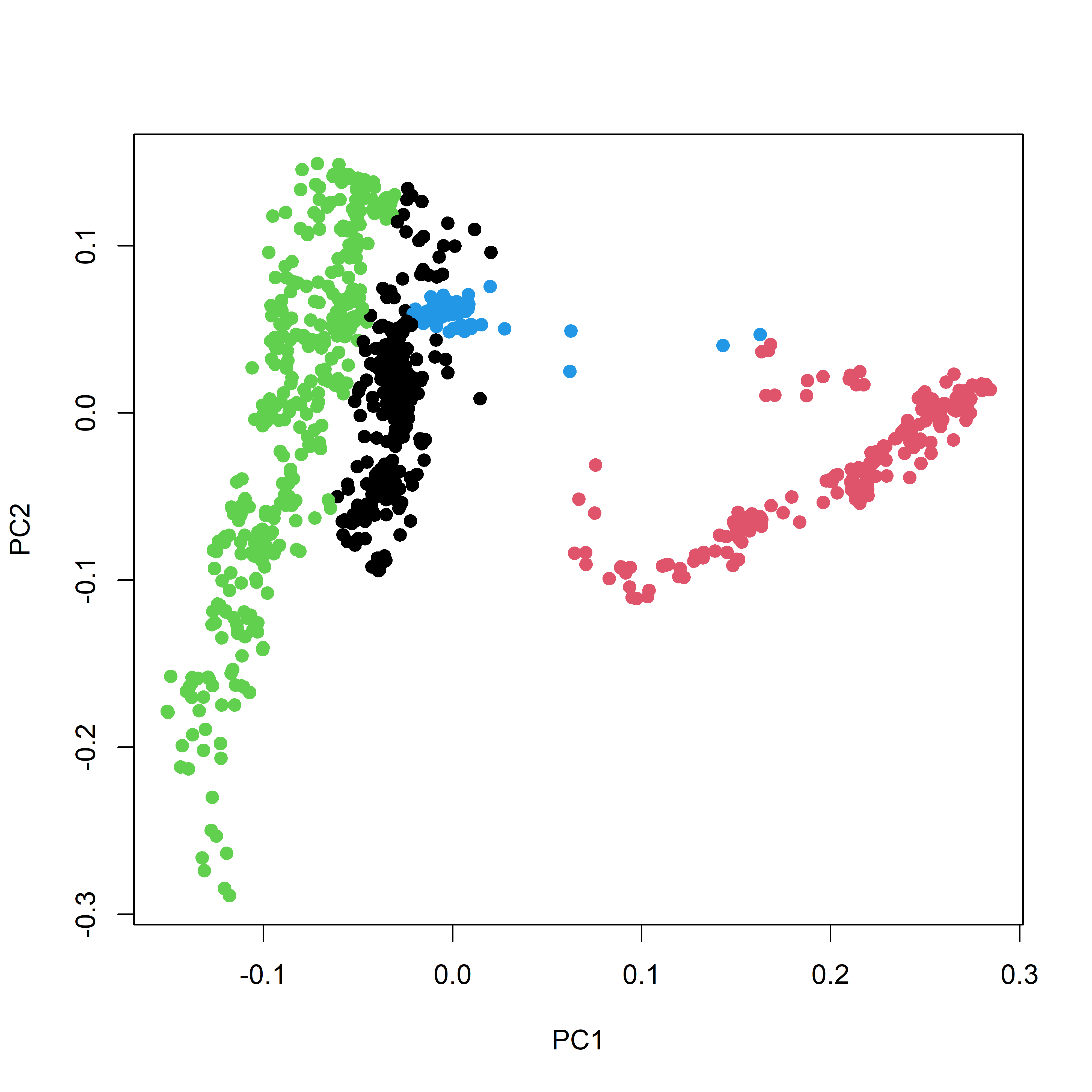}  &
\includegraphics[scale = 0.3]{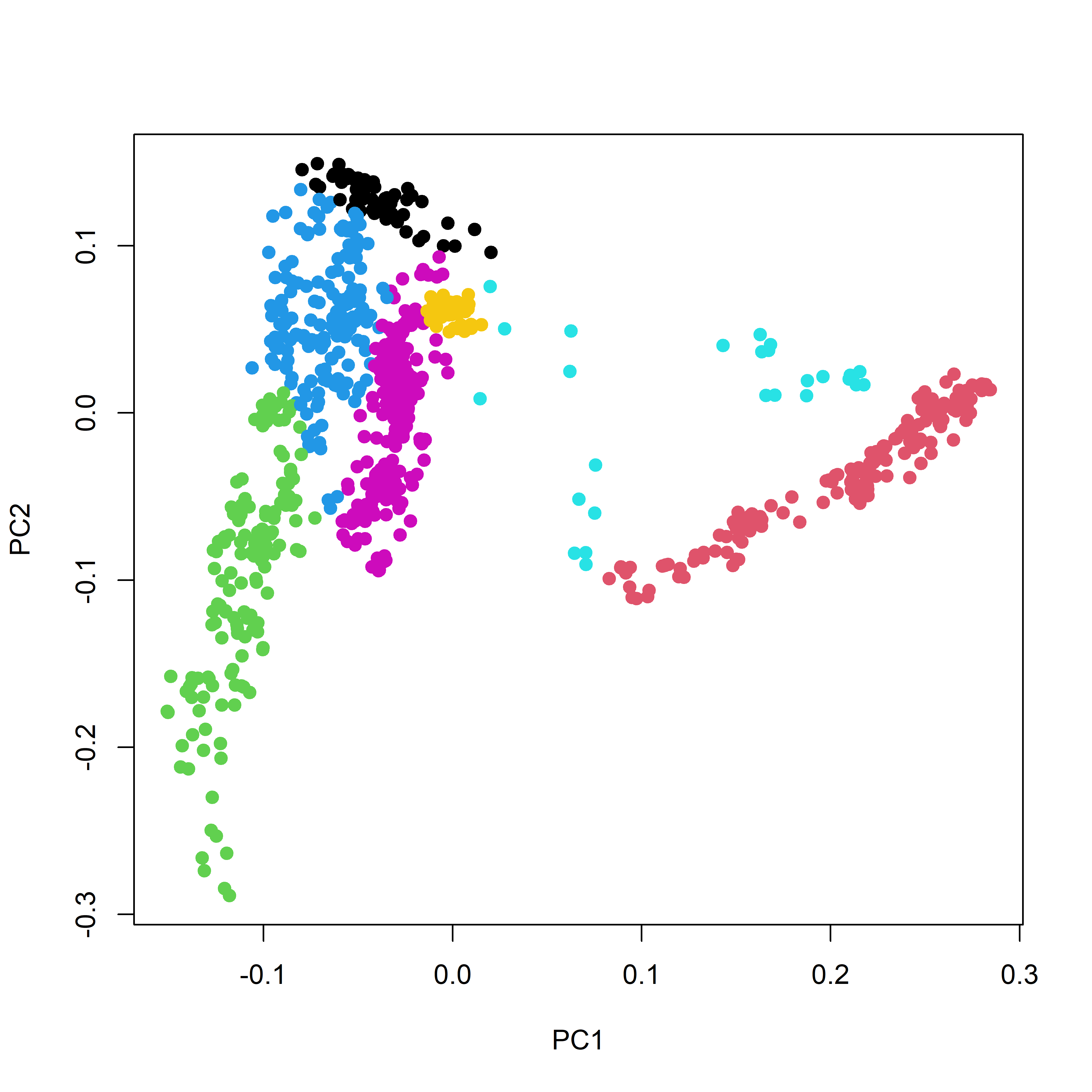} \\
(e) PCA plot of SESPC mixtures  &  (f) PCA plot of ESAG mixtures \\
\end{tabular}
\caption{Fiji region: ICL plots for the two mixtures and locations of the earthquakes coloured according to their estimated cluster. \label{fiji} }
\end{figure}

Since Fiji region contains more locations, we searched for the optimal number of components $K$, using $K=1,\ldots,10$. When fitting the SESPC, with parallel processing using 9 cores, the search required 69.98 seconds, whereas the ESAG required 110.83 seconds. Figures \ref{fiji}(a)--\ref{fiji}(b) present the ICL values for the mixtures of both distributions using the selected range of $K$. The SESPC mixtures selected 4 clusters, whereas the ESAG mixtures were less parsimonious, selecting 7 clusters. Figures \ref{fiji}(c)--\ref{fiji}(d) visualize, on the sphere, the clusters identified by each mixture of distributions. Figure \ref{fiji}(e)--\ref{fiji}(f) present the PCA projected data with colours identifying the different clusters. Looking at Figures \ref{fiji}(c)--\ref{fiji}(f) we can observe that the SESPC mixtures created clusters that are easier to separate in comparison to the clusters created using the ESAG mixtures. 

\begin{figure}[h!]
\centering
\begin{tabular}{cc}
\includegraphics[scale = 0.3]{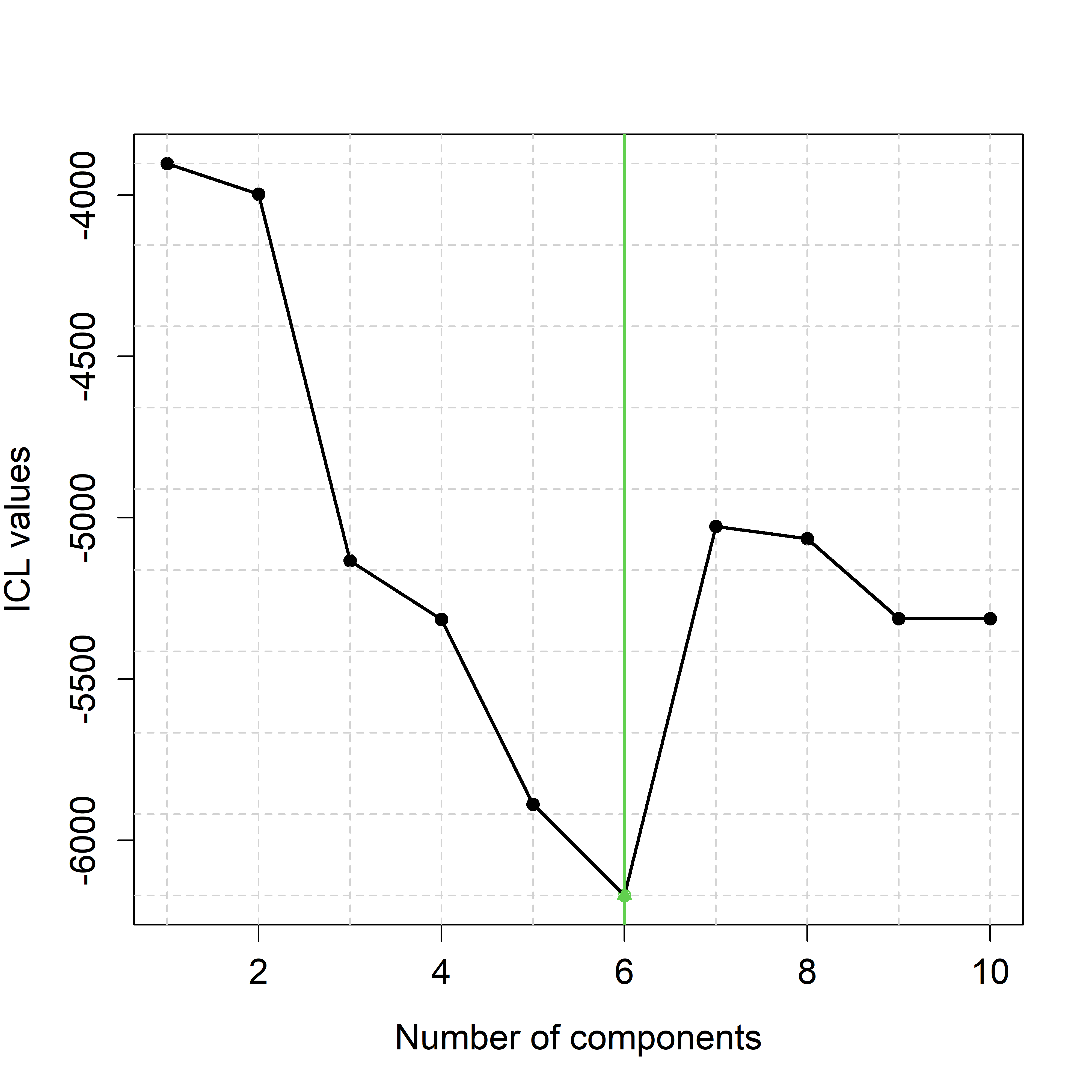} &
\includegraphics[scale = 0.27]{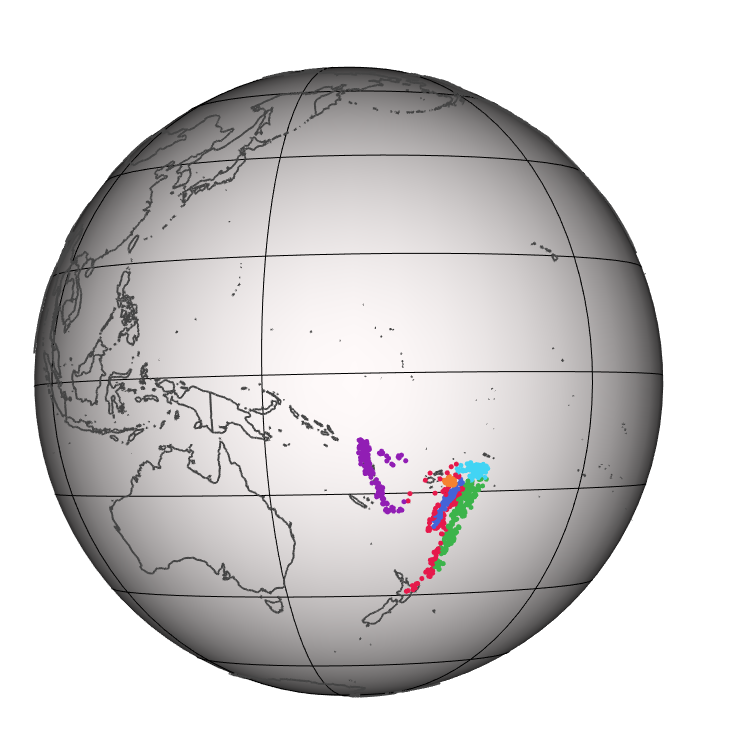} \\
(a) ICL plot &  (b) Spherical plot \\
\end{tabular}
\caption{Fiji region: ICL and spherical plot for the ESAG concomitant model. \label{fiji2} }
\end{figure}

We used again the magnitude of the earthquakes as a covariate in the concomitant model. The BIC and ICL of the SESPC mixtures reduced significantly, yet the concomitant model selected 4 clusters and the groups were affected, but to a small degree. The magnitude had a significant impact on the ESAG mixtures, where this time 6 clusters were selected. Figure \ref{fiji2} displays the ICL for the ESAG concomitant model and the spherical plot of the data with the clusters in different colours.

\subsection{Wine quality data}
The third dataset is the wine quality dataset\footnote{This dataset was downloaded from \href{https://archive.ics.uci.edu/dataset/186/wine+quality}{https://archive.ics.uci.edu/dataset/186/wine+quality}.}. 
The dataset contains 12 physicochemical features of red (1,599 observations) and white (4,898 observations) variants of the Portuguese "Vinho Verde" wine. For more details see \cite{cortez2009}.

\begin{figure}[h!]
\centering
\begin{tabular}{cc}
\includegraphics[scale = 0.55]{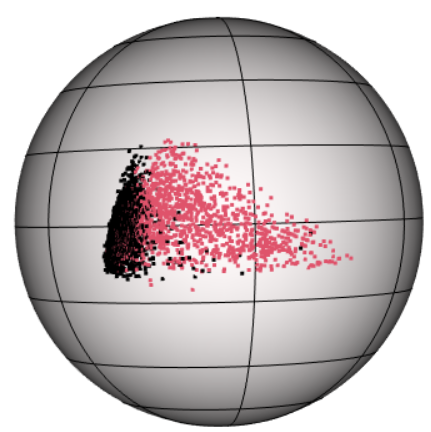} &
\includegraphics[scale = 0.55]{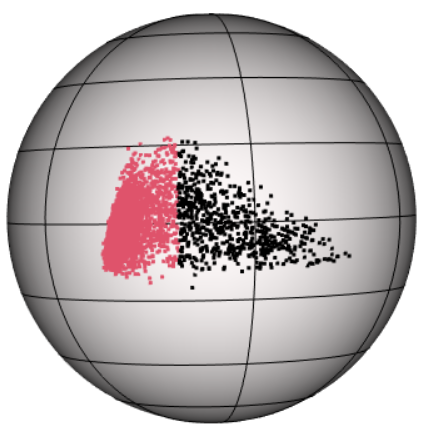} \\
(a) Sphere plot of the data & (b) SC mixture model \\
\includegraphics[scale = 0.55]{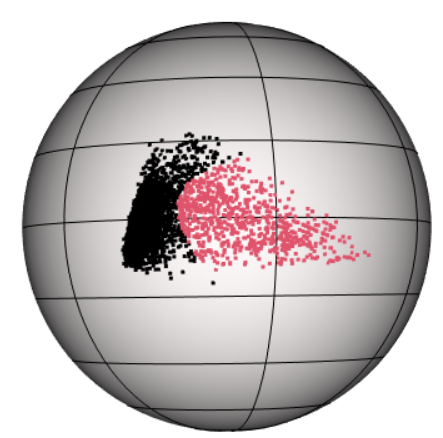} &
\includegraphics[scale = 0.55]{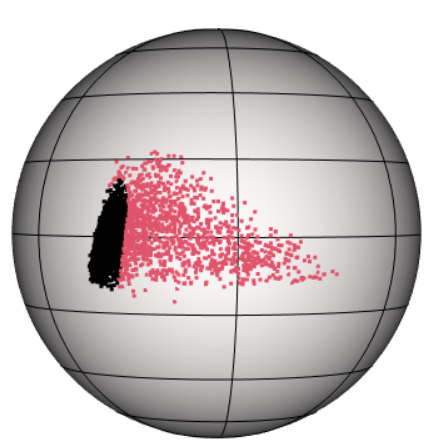} \\
(c) SESPC mixture model &  (d) ESAG mixture model 
\end{tabular}
\caption{Wine quality data: Sphere plot of the data with the true groups in different colours and the results of the SESPC and ESAG mixture models. \label{wine} }
\end{figure}

Figure \ref{wine}(a) displays the data projected onto the sphere with the 2 groups in different colours. The SESPC and ESAG mixture models both correctly identified 2 clusters and their estimated clusters are displayed in Figures \ref{wine}(c)--(d). The ESAG is shown to have identified the two clusters more properly and this is evident from the ARI values that were 0.546 and 0.700 for the SESPC and ESAG mixtures, respectively. The SC mixture model identified 13 clusters in the original data and 2 in the projected spherical data. However, the estimated clusters differ significantly as seen in Figure \ref{wine}(b) and this was evident in the ARI value that was the lowest, only 0.476.

\subsection{Wholesale data}
The fourth dataset refers to clients of a wholesale distributor\footnote{This dataset was downloaded from \href{https://archive.ics.uci.edu/dataset/292/wholesale+customers}{https://archive.ics.uci.edu/dataset/292/wholesale+customers}.}. It includes the annual spending in monetary units (m.u.) on 6 diverse product categories. There are two channels of distribution, either hotels/restaurants/catering or retail sales. 

\begin{figure}[h!]
\centering
\begin{tabular}{cc}
\includegraphics[scale = 0.55]{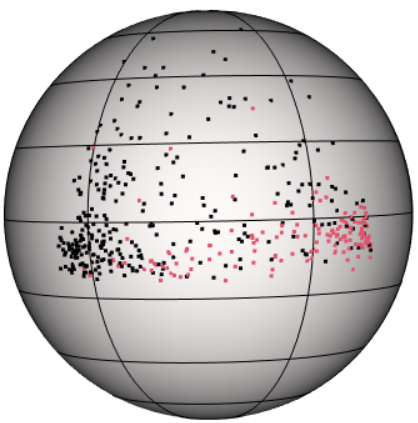} &
\includegraphics[scale = 0.55]{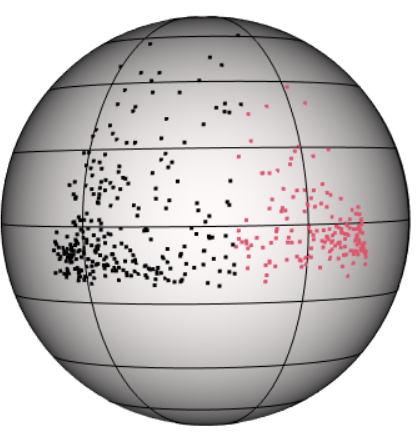} \\
(a) Sphere plot of the data & (b) SC mixture model \\
\includegraphics[scale = 0.55]{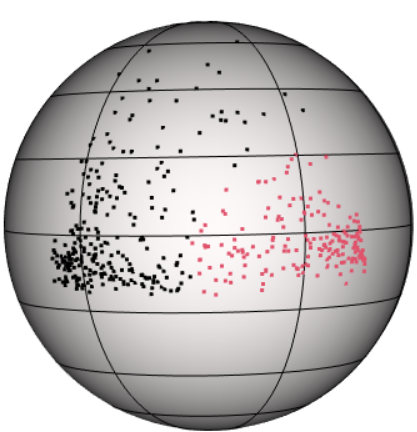} &
\includegraphics[scale = 0.55]{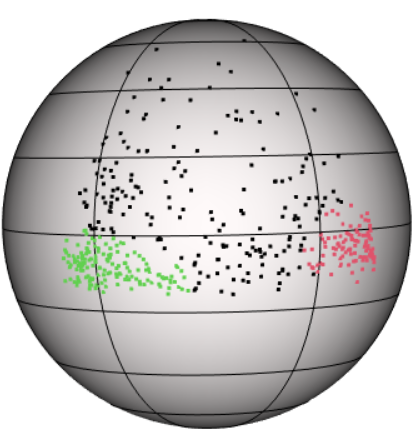} \\
(c) SESPC mixture model &  (d) ESAG mixture model 
\end{tabular}
\caption{Wholesale data: Sphere plot of the data with the true groups in different colours and the results of the SESPC and ESAG mixture models. \label{wholesale} }
\end{figure}

Figure \ref{wholesale}(a) displays the data projected onto the sphere with the 2 groups in different colours. The SC mixture model identified 14 clusters in the original data and 2 in the projected spherical data. Interestingly, the ESAG mixture model detected 3 clusters, whereas the SESPC model detected 2 clusters, as seen in Figures \ref{wholesale}(c)--(d). 

\section{Conclusions} \label{con}
We proposed mixture models based on two elliptically symmetric distributions defined on the sphere for clustering spherical data. The benefit of these mixture models over rotationally symmetric distributions is that they escape the isotropic covariance matrix assumption and allow for a more general covariance matrix that produces ellipsoidal contours. To address the dimensionality issues when it comes to fitting we proposed to project the hyper--spherical data onto the sphere and fit the proposed mixture models there. The functions to fit these models, along with other capabilities, are all included in the \textit{R} package \textsf{sphereclust} \citep{sphereclust2026}.

The simulation studies revealed that when the ground--truth model is the ESAG mixtures, both distributions perform equally well, but if the data are generated from a mixture of SESPC distributions, the SESPC mixture model outperforms the ESAG mixture model. Similar conclusions were derived for the hyper--spherical data cases.

When applied to two real spherical data containing the locations of earthquakes the two mixture models identified the same number of components in North America, albeit not clustering the same locations.  The clusters identified by the SESPC mixtures are easier to linearly separate than the ESAG mixtures. With Fiji region though, the SESPC mixtures identified 4 clusters, whereas the ESAG mixtures identified 7 clusters. The concomitant model made no impact on the North America dataset, whereas for the Fiji region it reduced the number of clusters identified by ESAG. When applied to two real hyper--spherical data the proposed mixtures applied to the projected data on the sphere performed better than mixtures of rotationally symmetric distributions applied either on the hyper--sphere or on the sphere. 

\clearpage
\bibliographystyle{apalike}
\bibliography{biblio}

\clearpage
\section*{Appendix}
\subsection*{Choice of the parameters regarding the hyper--sphere ESAG distribution}
\begin{figure}[h!]
\centering
\includegraphics[scale = 0.55]{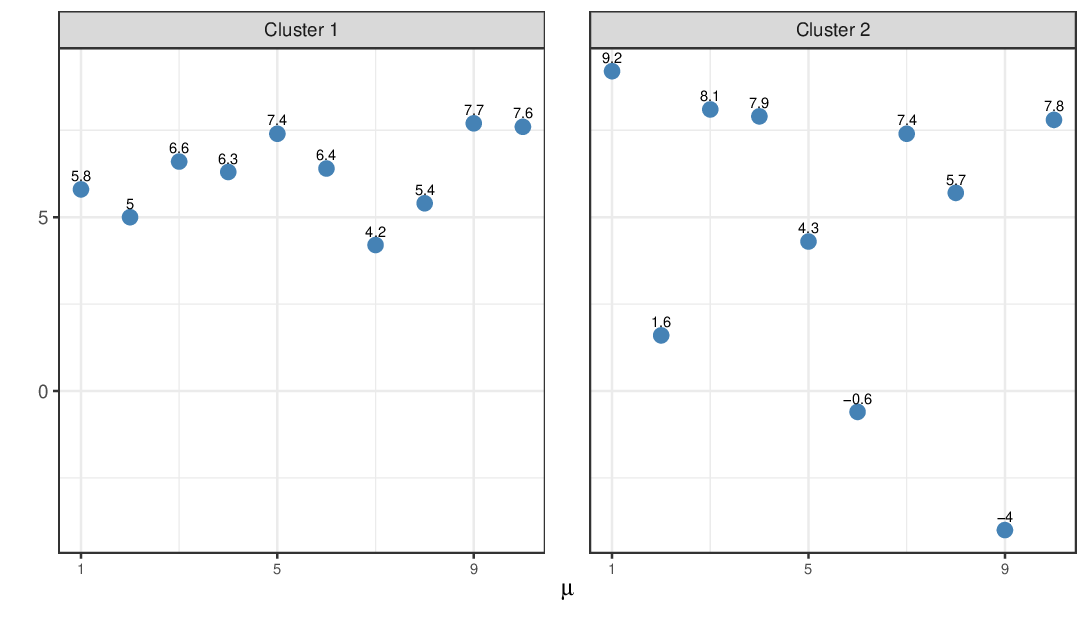} 
\caption{ Values of $\bm{\mu_1'}, \bm{\mu_2}'$ using $(\tau_1,\tau_2)=(20,20)$ for the 3 cases presented in Figure \ref{hyper_effect_of_gamma}.
\label{values_mu_hyper} }
\end{figure}

\begin{figure}[h!]
\centering
\begin{tabular}{c}
\includegraphics[scale = 0.55]{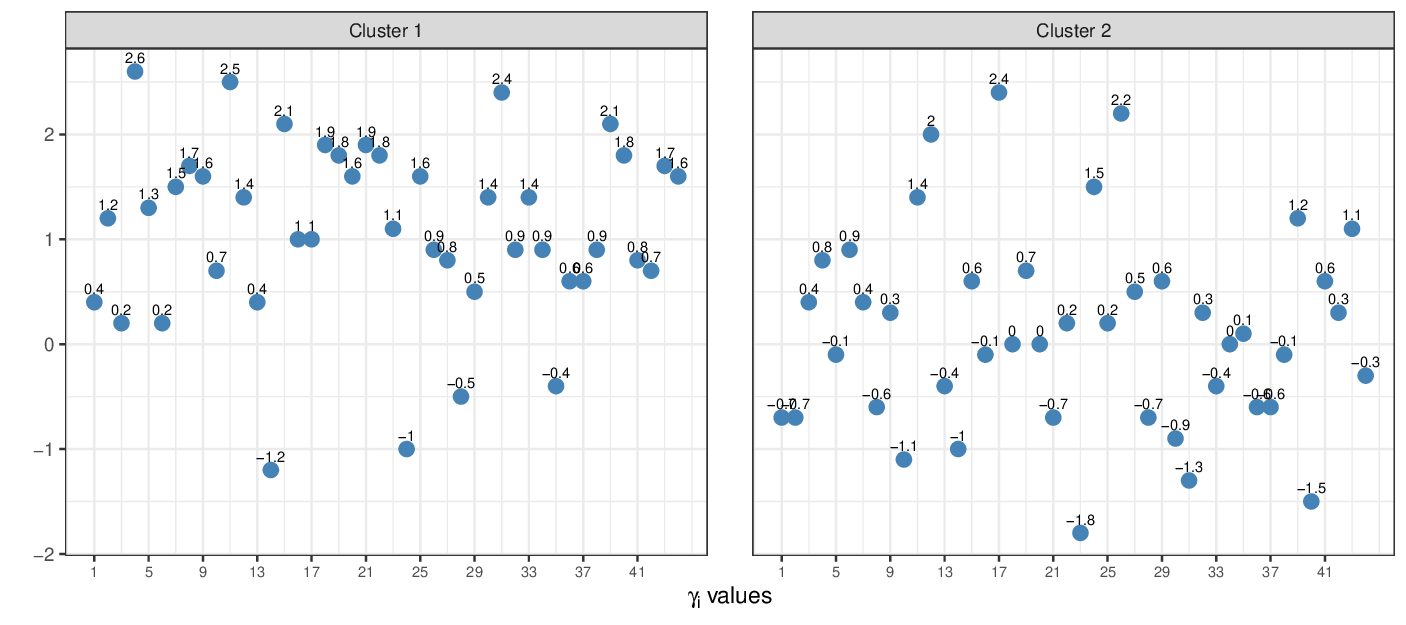} \\
(a) Case 1 \\
\includegraphics[scale = 0.55]{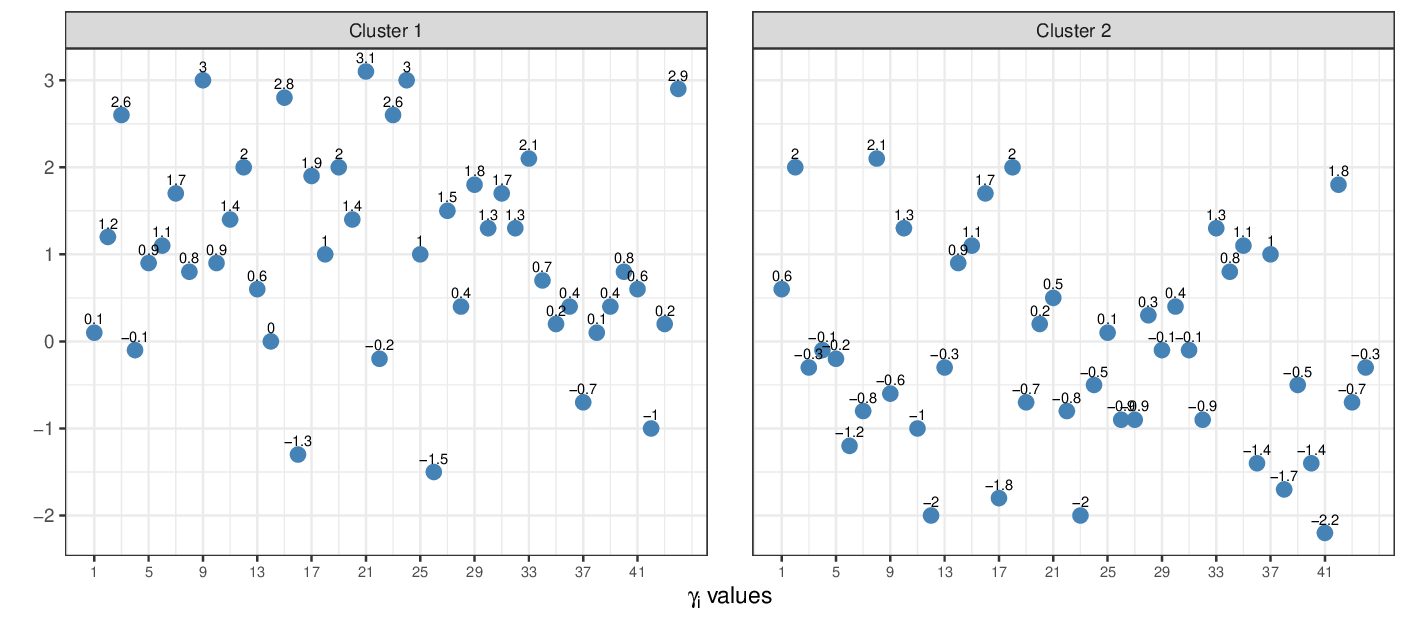} \\
(b) Case 2\\
\includegraphics[scale = 0.55]{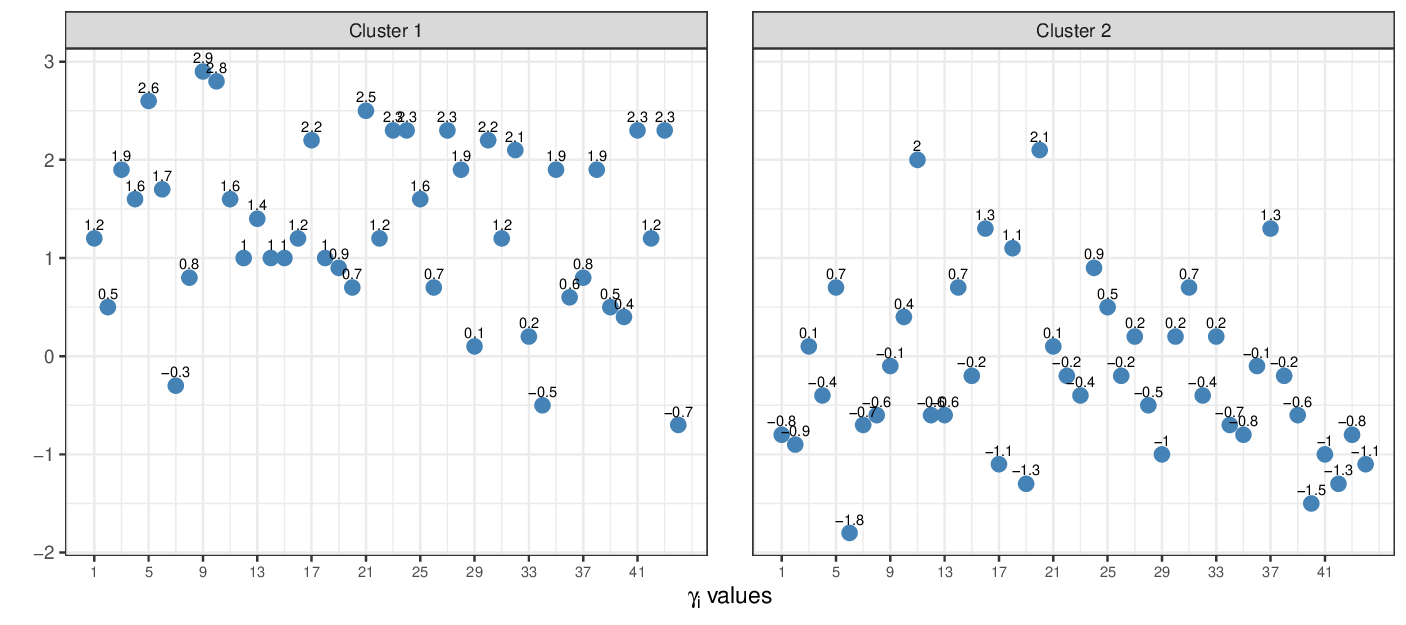} \\
(c) Case 3\\
\end{tabular}
\caption{Values of  $\bm{\gamma_1}, \bm{\gamma_2}$ for the 3 cases presented in Figure \ref{hyper_effect_of_gamma}.
\label{values_gam_hyper} }
\end{figure}

Since we are investigating a 10--dimensional ESAG distribution the vector $\bm{\gamma}$ consists of 44 values ($44=0.5(d^2 + d - 2)$). As a result, for each cluster, we consider the vectors $\bm{\gamma}_1, \bm{\gamma}_2$ by generating 44 random number from $N(0,1)$ for each case. In Figure \ref{values_gam_hyper}, the corresponding vectors $\bm{\gamma}_1, \bm{\gamma}_2$ are illustrated, for the three investigated cases presented in  Figure \ref{hyper_effect_of_gamma}. As for the choice of the mean vectors $\bm{\mu_1'}, \bm{\mu_2}'$, they were derived similarly to the  numerical analysis for Spherical data  and are presented in Figure \ref{values_mu_hyper}  using $(\tau_1,\tau_2)=(20,20)$. 

\end{document}